\documentclass{article}
\usepackage[T1]{fontenc}
\usepackage{geometry}
\usepackage[english]{babel}
\usepackage{array}
\usepackage{float}
\usepackage{booktabs}
\usepackage{amsmath}
\usepackage{amsthm}
\usepackage{amssymb}
\usepackage{graphicx}
\usepackage[authoryear,round]{natbib}
\usepackage[pdfusetitle,
 bookmarks=true,bookmarksnumbered=false,bookmarksopen=false,
 breaklinks=false,pdfborder={0 0 0},pdfborderstyle={},backref=false,colorlinks=false]
 {hyperref}

 \usepackage{todonotes}

\makeatletter

\providecommand{\tabularnewline}{\\}

\newenvironment{lyxcode}
	{\par\begin{list}{}{
		\setlength{\rightmargin}{\leftmargin}
		\setlength{\listparindent}{0pt}
		\raggedright
		\setlength{\itemsep}{0pt}
		\setlength{\parsep}{0pt}
		\normalfont\ttfamily}%
	 \item[]}
	{\end{list}}
\theoremstyle{remark}
\newtheorem{rem}{\protect\remarkname}
\theoremstyle{definition}
\newtheorem{defn}{\protect\definitionname}
\theoremstyle{plain}
\newtheorem{thm}{\protect\theoremname}

\newtheorem{assumption}{Assumption}

\makeatother

\providecommand{\definitionname}{Definition}
\providecommand{\remarkname}{Remark}
\providecommand{\theoremname}{Theorem}

\begin{document}

\title{Validation of machine learning based scenario generators}
\author{Gero Junike\thanks{Corresponding author. Institut für Mathematik, Carl von Ossietzky
Universität, 26129 Oldenburg, Germany, ORCID: 0000-0001-8686-2661,
E-mail: gero.junike@uol.de}, Solveig Flaig\thanks{Deutsche Rückversicherung AG, Kapitalanlage-Controlling, Hansaallee
177, 40549 Düsseldorf, Germany, E-Mail: solveig.flaig@deutscherueck.de}, Ralf Werner\thanks{Institut für Mathematik, Universität Augsburg, 86159 Augsburg, Germany,
ORCID: 0000-0001-8204-3757, E-mail: ralf.werner@math.uni-augsburg.de}}

\maketitle

\subsection*{\vspace{1cm}}

\begin{abstract}
Machine learning (ML) methods are becoming increasingly important in the design economic scenario generators for internal models. Validation of data-driven models differs from classical theory-based models. We discuss two novel aspects of such a validation: first, checking dependencies between risk factors and second, detecting unwanted memorization effects. The first task becomes necessary since in ML-based methods dependencies are no longer derived from a financial-mathematical theory but are driven by data. The need for the latter task arises since it cannot be ruled out that ML-based models merely reproduce the empirical data rather than generating new scenarios. To address the first issue, we propose to use an existing test from the literature. For the second issue, we introduce and discuss a novel memorization ratio. Numerical experiments based on real market data are included and an autoencoder-based scenario generator is validated with these two methods.
\end{abstract}

\subsection*{\vspace{2cm}}

\textbf{\emph{Keywords}} Nearest neighbor distance, economic scenario generator, memorization ratio,
Solvency 2, machine learning\textbf{\emph{}}\\
\textbf{\emph{}}\\
\textbf{\emph{JEL classification}} C14, C45, C63, G22

\subsection*{\vspace{2cm}}

\pagebreak{}
\begin{lyxcode}
\global\long\def\E{\mathbb{E}}%

\global\long\def\G{\mathbb{G}}%

\global\long\def\R{\mathbb{R}}%

\global\long\def\N{\mathbb{N}}%

\global\long\def\P{\mathbb{\mathbb{P}}}%
\end{lyxcode}

\section{Introduction}

Under Solvency 2, insurance companies are required to calculate a Solvency Capital Requirement, i.e., the value-at-risk at the 99.5\% level of their own funds over a one-year horizon. This is usually done by generating suitable scenarios using an economic scenario generator and taking the empirical 99.5\%-percentile of the loss. This paper focuses on the validation of such machine-learning based economic scenario generators.

Studies that
have looked at deriving a value-at-risk based on financial data using
neural networks, e.g., generative adversarial networks (GAN), include~\citet{tobjork2021value}, \citet{fiechtner2019risk} and \citet{flaig2022scenario}.  \cite{arian2022encoded} and \cite{buch2023estimating} use (variational) autoencoders  to estimate the value-at-risk. \citet{kondratyev2019market} propose a restricted Boltzmann machine for scenario generation.
Regulators in several countries, too, have authored papers that address
considerations associated with the use of ML methods in internal models.
For example, in 2019, the Nederlandsche Bank issued a paper discussing how artificial intelligence is currently used in finance;
see \citet{van2019generalAI}. There, the authors state that future
ML applications will ``use broader and better data to develop predictive
risk models'' and will ``increase the precision
of risk assessment\textquotedblright , see \citet[p. 28 and 29]{van2019generalAI}.
The German regulators, Bundesbank and BaFin, assert that ``the use of ML
methods can help to quantify risks more accurately and enhance process
quality, thereby improving financial firms' risk management''; nonetheless,
these authors also note that ``the limited transparency of the model's
behavior has consequences for the {[}...{]} model validation'', see
\citet[p. 3 and p. 7]{bafin2021ml}. Regulators in other countries, including
France (see \citet{dupont2020governance}) and the United Kingdom (see
\citet{jung2019machine}) have also issued papers on this topic.

One of the key processes for internal models in insurance companies
is validation, see \citet[Art. 124]{union2009directive}. \citet[p. 11]{bafin2021ml},
for example, notes that ``supervisors are focusing on any new or
much more pronounced risks that arise from ML methods.''
Therefore, it is important to determine what additional validation
measures need to be undertaken in order to prove the validity of a
given model when the scenario generator is ML-based.

Basically, in classical internal market risk models, the dependencies
between the risk factors (interest rates, equities, foreign exchange)
are modeled using financial mathematical tools such as correlations
or copulas, see \citet{sandstrom2016handbook} or \citet{pfeifer2018generating}.

In studies that have examined the use of ML methods to generate financial
scenarios, validation is carried out primarily by visual means or
using only a few statistical parameters, as is the case in \citet{wiese2019deep},
\citet{ni2020conditional} and \citet{wiese2020quant}. 
In a more quantitative way, \citet{cont2022tail}
propose an evaluation approach that focuses on the tail of the marginal distributions.

With ML-based models at least two new important issues have arisen which have to be addressed during the model validation process: 

i) Since in ML-based models the dependencies are not explicitly modeled, an additional validation has to be performed  to show that the dependencies (or, more generally, the multivariate distribution) generated by the model resemble the empirical multivariate distribution including the empirical dependencies. 

ii) Another issue associated with generative ML models is the
so-called \emph{memorization effect}, see for instance~\citet{nagarajan2018theoretical}.
In this phenomenon, instead of generating new scenarios for risk management
purposes, the internal model simply replicates the empirical data
used in the learning process. It is well known that neural networks
can memorize some training data, see \citet{wei2024memorization}
and references therein. \citet{arora2017gans}, for example,
state that one of the most obvious pitfalls associated with a GAN
training is that a GAN may simply
memorize the training data. A similar statement holds true for autoencoders
as well, see \citep{radhakrishnan2018memorization}. GANs and autoencoders usually consist of two (potentially coupled) neural networks and both types have already been applied as economic scenario generators in the literature (see above). 

The literature on estimating the memorization effect in generative methods focuses mainly on large language models, see \cite{wei2024memorization} and image generation, see \citet{borji2019pros} and \citet{borji2022pros} for survey papers. In particular, \citet{lopez2016revisiting} and \citet{xu2018empirical} propose a 1-nearest neighbor classifier to measure memorization. \citet{bai2021training} propose a memorization-informed Fréchet inception distance with memorization penalty to evaluate GANs based on cosine similarity. However, the proposed measures of the memorization effect are mainly heuristically motivated metrics which focus on image generation without analyzing the behavior as the sample size of the data increases. \citet{meehan2020non} and \citet{van2021memorization} propose a probabilistic approach to measure memorization. However, it requires some data to be withheld for validation, which means that the amount of training data is reduced. This is particularly unfavorable in applications where the training data is sparse.

In the following, we suggest two approaches how to deal with issues i) and ii):
To address point i), we propose to use a test based on nearest neighbor distances, as studied by \citet{weiss1960two,bickel1983sums,schilling1986multivariate,henze1988multivariate,mondal2015high}
and \citet{ebner2018multivariate}, which cover both dependencies and marginal distributions at the same time. 
To address point ii), we develop a new measure called \emph{memorization ratio} to capture the memorization effect.  
We classify an empirical data point as memorized if a generated data point lies in an unusually small neighborhood around that empirical data point. 
Our memorization ratio is easy to implement and to interpret. 
We show that the memorization ratio convergences to a dimension-independent constant when the amount of data increases. 

We test both measures on simulated and real market data. In particular, our experiments include a real-world example: we use an autoencoder to generate new interest rate scenarios in a high-dimensional space. The training data consists of a term structure interest rate time series from 2000 to 2022. The autoencoder is validated using a nearest neighbor test and our new menorization ratio, demonstrating that the memorization ratio is indeed effective. We further provide in-sample and out-of-sample validation for our data sets. 

In particular, out-of-sample validation is a common technique in machine learning to estimate the generalization error, see for instance \citet[Sec. 8]{GoodBengCour16}, or \cite{gu2021autoencoder} who apply out-of-sample validation for autoencoders in a financial context. 

The remaining paper is structured as follows: 
In Section \ref{Sol2}, we provide a short background on risk calculation and validation under Solvency 2. 
In Section \ref{NN} we give a brief introduction to neural networks and autoencoders. 
In Section \ref{sec:GAN-evaluation-measures}, we introduce the two proposed validation techniques and  in Section \ref{NumerExp}, we show how these techniques can be used for model validation. 
We consider a one-dimensional real market data set (S\&P 500) for illustrative purposes.
We also work with multidimensional, autocorrelated simulated data and consider an autoencoder trained on real market data to generate a high-dimensional interest rate term structure.
Section \ref{conc} concludes.

\section{Background: Risk calculation and model validation under Solvency 2}\label{Sol2}

Solvency 2 allows insurers to calculate their risk, which must be backed by capital, in two different ways. 
Either the standard model is used for this purpose or the insurer develops its own internal model.
While in the standard model the method and the weighting factors for all risks are fixed, in the second case the insurer is free to choose the methods, which are quite often based on economic scenarios. 
More details on the standard model are for example given in \citet[Article 112]{union2009directive} or \citet[Appendix I]{sandstrom2016handbook}. 
For a critical review of the standard model let us refer to \cite{scherer2021standard}.  
An introduction to economic scenario generators is for instance provided by \cite{pedersen2016economic}. See \cite{begin2021complex} for Monte Carlo simulations for internal models. \cite{varnell2011economic} provides a non-technical view on economic scenario generation under Solvency 2. 
A comparison of the standard model and internal models for calculating Solvency 2 capital requirements is carried out in \cite{shedari2016solvency} or \cite{gatzert2012quantifying}.

Often, a historical simulation is implemented in order to estimate a (short horizon) value-at-risk, see \cite{hendricks1996evaluation} and \cite{beder1995var} for an analysis of the empirical performance of such simulation. 
In a historical simulation, similar to bootstrapping, empirical data is simply copied to ``generate'' future scenarios. 
Similarly, \cite{demirel2002generation}, \cite{muller2004bootstrapping} and \cite{albeanu2008using} propose a bootstrap method for scenario generation. 
A critical view on the topic is for example due to \cite{pritsker2006hidden}.
Quite related, \cite{adesi2014simulating} propose a filtered historical simulation approach, where historical returns are multiplied by the ratio of the historical volatility and the current volatility.

To ensure comparability of risk calculations, insurers determining
their risk using an internal model must meet certain requirements.
A very important one relates to the regular validation of the internal
model. According to \citet[Art. 124]{union2009directive}, this ``includes
monitoring the performance of the internal model, reviewing the ongoing
appropriateness of its specification, and testing its results against
experience''.

Current internal models for market and non-life underwriting risk
often use Monte-Carlo simulation techniques to derive the risk of
the (sub)modules and then use correlations or copulas for aggregation,
see \citet{pfeifer2018generating}.
On the market risk side, the risk calculation is often based on an
economic scenario generator (ESG); this approach produces realistic
scenarios how the risk factors may evolve over a certain time horizon, which is set to one year in Solvency 2,
see \citet[p. 190]{cadoni2014internal}. 
Traditional ESG approaches implement
financial-mathematical models for all relevant risk factors (e.g.
interest rate, equity) and their dependencies. Under these scenarios,
the asset and liability portfolio of the insurer is evaluated
and the risk is given by the 99.5\%-percentile of the loss in these
scenarios. The models for non-life underwriting risk also simulate
scenarios as to how the claims evolve and these scenarios are then
used to derive the risk, see \citet[p. 192]{cadoni2014internal}.

However, it is not appropriate to limit the validation of ML-based
models to the methods used for classical financial-mathematical models.
In this context, the German regulator, BaFin, warns that ``Supervisory
practice for ML methods can {[}...{]} be derived from the existing
framework. At the same time, an outlier analysis, also supported by
this consultation, is currently surveying the areas in which the supervisory
inspection approach needs to be fleshed out in order to cater to the
peculiarities of using ML methods.'', see \citet[p. 11]{bafin2021ml}.
This means that the validation for an internal model can basically
remain the same, but any special features of the ML-based approach
must additionally be taken into account and validated.

\section{Background: Neural networks and autoencoders}\label{NN}

\begin{figure}[H]
\begin{centering}
\includegraphics[scale=0.5]{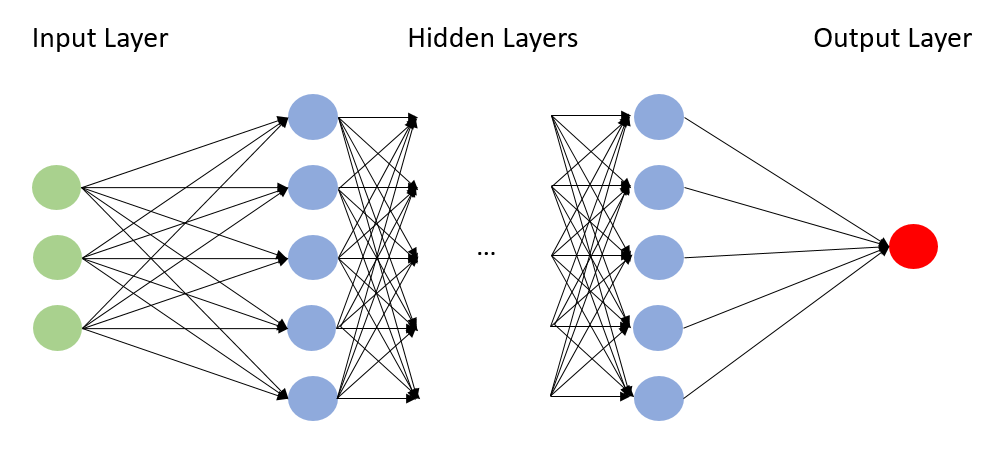}
\par\end{centering}
\caption{\protect\label{fig:NN}Neural network structure, based
on \citet[Fig. 1]{fernandez2021neural}.}
\end{figure}

A \emph{neural network} is a system that is broadly inspired from the human brain. 
It can transform a (one- or more-dimensional) input into a (one- or more-dimensional) output. 
A neural network consists of \emph{neurons} which are arranged in \emph{layers}. 
The neurons of one layer are connected to the neurons of the next layer. All starts
with the \emph{input layer}, which is then followed by one or more
so called \emph{hidden layers} and ends with the \emph{output layer}, as Figure \ref{fig:NN}
based on \citet[Fig. 1]{fernandez2021neural} illustrates. A detailed and more thorough treatment of neural networks can be found in \cite{GoodBengCour16}.

All neurons transform their multi-dimensional input (which comes from
either the input layer or the proceeding hidden layer) via a so called
\emph{activation function} into an one-dimensional output which is
then transferred to all the neurons in the next layer.

A typical choice for an activation function $g$ is the rectified linear unit $g(t)=\max(t,0)$. For a neuron input $x\in\mathbb{R}^{n}$ and \emph{parameters} or \emph{weights} $w\in\mathbb{R}^{n+1}$ the output of the neuron is then given by $g\big(\sum_{i=1}^{n}w_{i}x_{i}+w_{n+1}\big)$.

The parameters in the activation function in each neuron are often initialized randomly. 
Afterwards, these parameters are optimized when
giving the neural network training data consisting of pairs of input
data and the desired output so that the output of the neural network
ressembles the desired output given the training data. 

In the numerical section, we will use a special type of neural network, an autoencoder, in a generative fashion as an economic scenario generator. 
For this purpose, let us briefly explain the main functionality of an autoencoder.
An \emph{autoencoder} consists of two neural networks: an \emph{encoder}
network $\mathcal{E}_{\phi}:\mathbb{R}^{d}\to\mathbb{R}^{k}$ with
parameters $\phi$ and a \emph{decoder }network $\mathcal{D}_{\theta}:\mathbb{R}^{k}\to\mathbb{R}^{d}$
with parameters $\theta$. 
Autoencoders can be employed for a variety of tasks; most prominently, autoencoders can be used for dimension reduction, where $d \gg k$ holds.
In such a setup, autoencoders can be interpreted as a nonlinear version of the traditional principal component analysis.
For more details on autoencoders as well as on further applications, let us refer to \citet{GoodBengCour16}.

Concerning the training of autoencoders, let $E$ be a $d$-dimensional random variable and let $E_{1},...,E_{M}$ be $M$ independent copies of $E$. 
Since the decoding and encoding is not perfect, a reconstruction error $E_{m}-\mathcal{D}_{\theta}\big(\mathcal{E}_{\phi}(E_{m})\big)$ on the $m$-th input $E_m$ cannot be avoided.
Training is usually achieved by minimization of these reconstruction errors by an appropriate reconstruction loss function (like the following mean-square loss $L$, or more generally some $p$-norm) via some (stochastic) gradient descent method: 
\[
L(\phi,\theta):=\frac{1}{M}\sum_{m=1}^{M}\left\Vert E_{m}-Y_m\right\Vert^2 _{2} \longrightarrow \min! \quad\quad \text{with } Y_m := \mathcal{D}_{\theta}\big(\mathcal{E}_{\phi}(E_{m})\big).
\]
Let us assume that the parameters $\phi$ and $\theta$ of the autoencoder are obtained by some training method and that (in-sample and out-of-sample) reconstruction errors are small enough.
Then, the decoder can be used in a generative fashion, if the (low-dimensional) distribution of the so-called \emph{latent factors} $Z := \mathcal{E}_{\phi}(E)$ is (approximately) known:
to generate $N$ new samples for $E$, let $Z_{1},...,Z_{N}$ be independent copies of $Z$ and set
\[
G_{n}:=\mathcal{D}_{\theta}(Z_{n}),\quad n=1,...,N.
\]
In our selected example in Section \ref{subsec:Auto-Encoders-and-interest}, we will see that although $Z$ might not be multivariate normal distributed, it can still be approximated sufficiently well by such a distribution, which allows to use the autoencoder in a generative manner as an economic scenario generator. 

If $Z = \mathcal{E}_{\phi}(E)$ cannot be easily approximated by some appropriate multivariate distribution, the autoencoder can be enhanced to an \emph{adversarial} or to a \emph{variational autoencoder}.
Very briefly speaking, these enhanced setups change the encoder-decoder pair $(\mathcal{E}_{\phi}, \mathcal{D}_{\theta})$ in such a way that $\mathcal{E}_{\phi}(E)$ is approximately multivariate normal distributed while keeping the encoding-decoding performance, see~\citet{KingmaWelling2019} and~\citet{MakhzaniSJG15} for more details.

\section{Validation techniques for machine learning based scenario generators}\label{sec:GAN-evaluation-measures}

During the usual validation cycle, insurers already investigate a variety of statistical properties of a scenario generator. 
For example, stability and sensitivities are typically already part of established validation procedures.
However, within an ML-based approach, some additional validation is required as we have already pointed out in the introduction:
Since the multivariate distribution of the risk factors is not derived from a financial mathematical model, it has to be verified that the generator only produces samples from the same (or at least very similar) distribution as implied by the data sample.
Further, since a pure resampling method is not intended by an ML-based generator, it is necessary to analyze for potential memorization effects.

Here, we suggest a purely data-driven non-parametric evaluation approach, because the non-parametric nature is one of the main advantages of using ML-based generators.
Such methods which can be applied to any model are generally referred to as \emph{model agnostic, }see e.g., \citet[p. 4]{borji2019pros}.
Furthermore, we prefer quantitative measures which are easily interpretable and computationally not too expensive, thus fostering their use in validation purposes in a Solvency 2 regime. 
For this purpose, we split the recommended validation techniques into two components:
\begin{enumerate}
\item[(1)] the alignment of the multivariate distributions in the given empirical data and in the generated data, and
\item[(2)] the memorization effect.
\end{enumerate}
Measures for these two categories are discussed in the subsequent Subsections \ref{subsec:NNC}
and \ref{subsec:MR}.
Like any other test statistics, these kind of methods could be used not only for validation, but also for hyperparameter optimization of the ML models.
Of course, additional validation, for example of marginal tail behavior, could be carried out using the
quantitative measures based on value-at-risk and expected shortfall as proposed in \citet[p. 21]{cont2022tail} or by comparison to benchmark portfolios as in~\citet[Chapter 3]{flaig2022scenario}. 

For the mathematical discussion in the remainder of this paper, we rely on the following notation:
Let $(\Omega,\mathcal{F},\ensuremath{P)}$ be a probability space and $d\in\mathbb{N}$ the number of risk factors that are modeled. 
Further, let $E:\Omega\rightarrow\R^{d}$ denote a random vector describing the empirical data.
The data generated by some ML-based generator is denoted by the random vector $G:\Omega\rightarrow\R^{d}$. 
Finally, let $E_{1},...,E_{M}$, and $G_{1},...,G_{N}$ be independent copies of the random vector $E$ and $G$, respectively. 
Since the consideration of memorization effects is mainly important for continuous distributions, we make the following main assumption:
\begin{assumption}\label{H0} The random variables $E_{1},...,E_{M},G_{1},...,G_{N}$
are independent and identically distributed and have a piecewise continuous
and bounded probability density function.
\end{assumption}

\subsection{Measure for the alignment of the multivariate distributions: Nearest neighbor coincidence}\label{subsec:NNC}

The test on the alignment of the multivariate distribution including dependency and marginal distributions of the empirical and the generated data can be reformulated as a multivariate two-sample test where we want to measure the equality of two multivariate distributions based on two sets of independent observations. 
For this purpose, and in alignment with the above considerations, we prefer the \emph{nearest neighbor coincidence measure} as defined by \citet{schilling1986multivariate} and as further developed by \citet{mondal2015high}. 
This test statistic is rather easy to interpret and to implement, computationally cheaper than alternative methods (for example, based on optimal transport) and the numerical experiments by \citet{mondal2015high} are very promising.
Of course, alternative possibilities for checking equality of distributions are available, e.g., maximum
mean discrepancy, average log-likelihood, F1-score, and many more, see for instance \citet{borji2019pros}.

For the definition of the nearest neighbor coincidence let us consider the indicator function $\delta_{E_{m}}(r)$ which takes the value 1 if the $r$-nearest neighbor (measured by the Euclidean distance) of $E_{m}$ within the set $\{E_{1},...,E_{M},G_{1},...,G_{N}\} \backslash \{E_m\}$ is an empirical data point and 0 if it is a generated data point. 
Accordingly, the indicator function $\delta_{G_{n}}(r)$ is 1 if the $r$-nearest neighbor is a generated data point and 0 otherwise. 
As in \citet{mondal2015high}, we define for $k\in\mathbb{N}$:
\begin{equation}
T_{NN1,k}=\frac{M|T_{E,k}-\frac{M-1}{N+M-1}|+N|T_{G,k}-\frac{N-1}{M+N-1}|}{N+M},\label{eq:mondal}
\end{equation}
where
\[
T_{E,k}=\frac{1}{Mk}\sum_{m=1}^{M}\sum_{r=1}^{k}\delta_{E_{m}}(r)\quad\text{and}\quad T_{G,k}=\frac{1}{Nk}\sum_{n=1}^{N}\sum_{r=1}^{k}\delta_{G_{n}}(r).
\]
The test statistic $T_{E,k}$ (the test statistic $T_{G,k}$) counts for each empirical (generated) data point the number $r\leq k$ of neighbors which are also empirical (generated) data points and takes the average.
Under Assumption \ref{H0}, $T_{NN1,k}$ should be close to 0. 
If the empirical data points and the generated data points are rather distinct from each other (i.e.\ stem from different distributions), $T_{NN1,k}$ should take larger values due to lack of full mixing of the two samples. 
In particular, for $M=N$ and if the generated data is far off the empirical data, $T_{E,k}$ and $T_{G,k}$ should be both close to their maximum value 1 and $T_{NN1,k}$ should be close to
$\frac{1}{2}$.
\begin{rem}
\citet{mondal2015high} prove that $\sqrt{N+M}\cdot T_{NN1,k}$ is asymptotically distributed as a sum of two correlated half normals if Assumption \ref{H0} holds and if $N$ and $M$ grow to infinity with $\frac{M}{N}\to\alpha$ for some $\alpha>0$. 
It is further shown that under Assumption \ref{H0}, the expectation of $T_{E,k}$ is equal to $\frac{M-1}{N+M-1}$ and the expectation of $T_{G,k}$ is equal to $\frac{N-1}{M+N-1}$.
\end{rem}
\begin{rem}
To compute $T_{NN1,k}$, we have to find the minimum entry of a vector of length $M+N$ exactly $M$ respectively $N$ times, to obtain $T_{E,k}$, resp.\ $T_{G,k}$. 
The computational effort for the computation of $T_{NN1,k}$ is therefore at most $O\left((N+M)^{2}\right)$.
\end{rem}
\noindent
An implementation of $T_{NN1,k}$ can be found in Appendix \ref{sec:R-Code}.

\subsection{Measure for the detection of memorization: memorization ratio}\label{subsec:MR}

The statistic $T_{NN1,k}$ will lead to very good scores if the generated data are drawn with repetition from the empirical ones (bootstrapping). 
From a risk management perspective, this is not the optimal result because the main intention of employing a generative ML-based generator is to create new scenarios that could happen rather than memorizing scenarios that have actually taken place, see \citet[p. 2]{chen2018bayesian}.
\citet[p. 1]{bai2021training} emphasize that ``unintentional memorization is a serious and common issue in popular generative models''. 
Therefore, we need some (preferably interpretable) quantitative measure to detect whether generated data points match empirical ones or not.

\begin{figure}[H]
\begin{centering}
\includegraphics[scale=0.3]{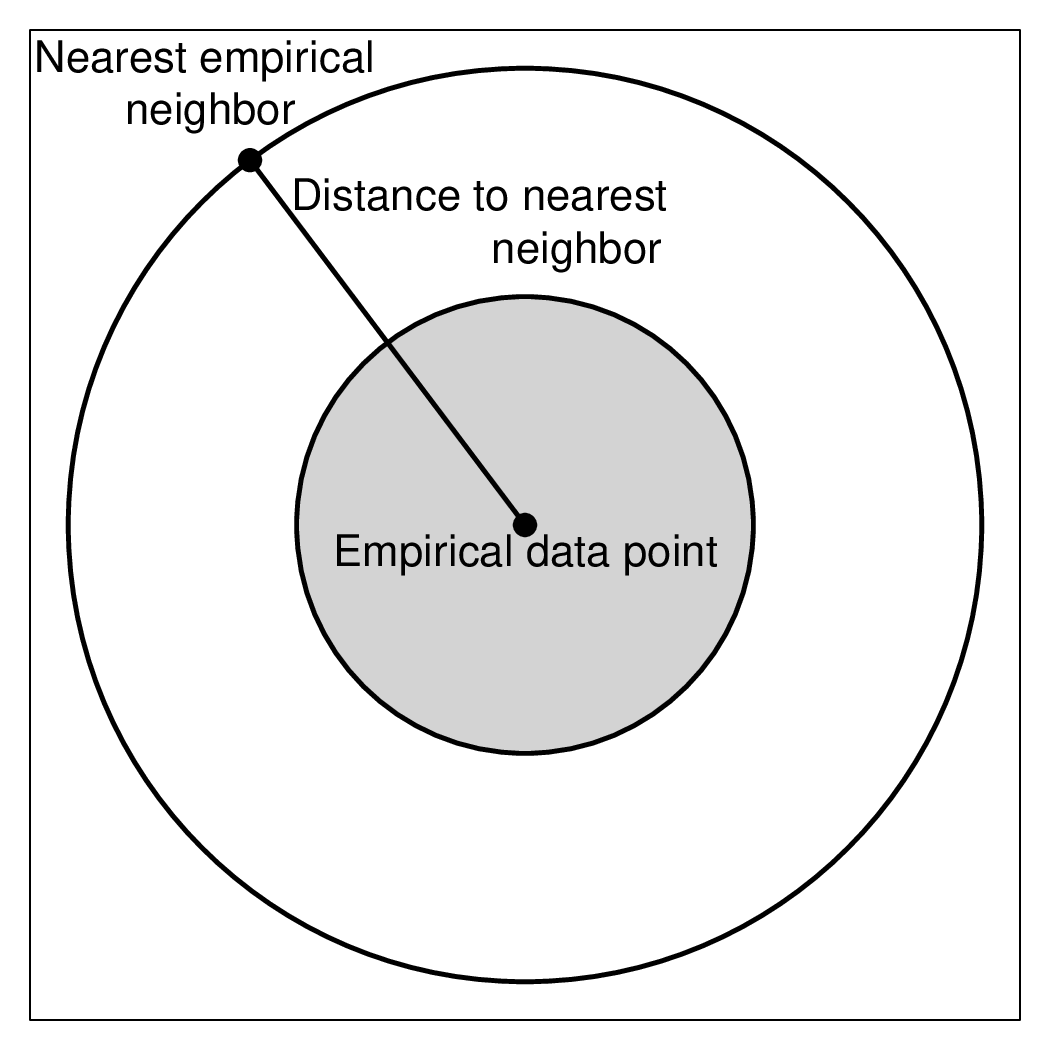}
\par\end{centering}
\caption{\protect\label{fig:NN25}Memorization ratio with $\rho=0.25$. If there
is at least one generated point within the gray area, the empirical
data point is considered memorized. The gray area corresponds to $25\%$
of the whole area of the circle.}
\end{figure}

In a multi-dimensional space, it is highly unlikely for generated and empirical data points to match exactly, so we classify an empirical data point as being memorized if a generated data point lies in ``an unusual small neighborhood'' around this empirical data point. 
\begin{defn}
\label{def:MR}Let $\rho\in(0,1]$. The \emph{memorization ratio $\Pi_{M,N}^{\rho}$}
is defined as: 
\begin{equation}
\Pi_{M,N}^{\rho}=\frac{1}{M}\sum_{m=1}^{M}\mathbf{1}_{[0,\,\rho^{\frac{1}{d}}R_{m})}\Big(\min_{n=1,...,N}\:\left\Vert G_{n}-E_{m}\right\Vert _{2}\Big)
\end{equation}
where 
\begin{equation}
R_{m}:=\min_{m^{\prime}\neq m}\left\Vert E_{m}-E_{m^{\prime}}\right\Vert _{2},\quad m=1,...,M.\label{eq:Rm}
\end{equation}
\end{defn}
The random variable $R_{m}$ denotes the distance between $E_{m}$ and its nearest empirical neighbor. 
An Euclidean ball of radius $\rho^{\frac{1}{d}}R_{m}$ has a volume equal to a fraction $\rho$ of the volume of a ball of radius $R_{m}$. 
An empirical data point $E_{m}$ is considered \emph{memorized} if there is at least one generated data point inside the ball of radius $\rho^{\frac{1}{d}}R_{m}$ around $E_{m}$, compare with Figure \ref{fig:NN25}.
The memorization ratio equals the percentage of elements $E_{1},...,E_{M}$ which are memorized. 
If all empirical data is memorized, then $\Pi_{M,N}^{\rho}=1$. 
If no empirical data is memorized, then $\Pi_{M,N}^{\rho}=0$. 
An implementation of $\Pi_{M,N}^{\rho}$ can be found in Appendix \ref{sec:R-Code}.
\begin{rem}\label{OnRho}
The parameter $\rho$ allows to control the interpretation of memorization:
The smaller $\rho$, the less likely it is that an empirical data
point is deemed memorized. In the extreme case that $\rho$ approaches
zero, an empirical data point will only be considered memorized if
there is an identical generated data point. This happens, for example,
when the empirical data is simply copied, e.g., during bootstrapping.
The greater $\rho$, the more likely it is that an empirical data
point is considered memorized. The special case $\rho=1$ is related to the measure ``1-NN accuray (real)'' in \citet[p. 3]{xu2018empirical}. 
Most importantly, our approach is more general and backed by mathematical results: for instance, the subsequent Theorem \ref{thm:MR_Konvergenz} establishes convergence of the memorization ratio $\Pi_{M,N}^{\rho}$ to a dimension-independent limit.
\end{rem}
\begin{thm}
\label{thm:MR_Konvergenz}Let $\alpha>0$ and $\rho\in(0,1]$. Under
Assumption \ref{H0}, it holds in the mean square sense that
\[
\lim_{M, N \to\infty,\frac{M}{N}=\alpha}\Pi_{M,N}^{\rho}=\frac{\rho}{\rho+\alpha}.
\]
\end{thm}
\begin{proof}
The proof can be found in Appendix \ref{sec:Appendix:-Proof-of}.
\end{proof}
\begin{rem}
According to Theorem \ref{lem:weiss} of \citet{weiss1960two}, the limit $\frac{\rho}{\rho+\alpha}$ of the memorization ratio under \ref{H0} equals the asymptotic probability that an arbitrarily empirical
data point is memorized.
\end{rem}

The convergence of the memorization ratio to its theoretical limit stated in Theorem \ref{thm:MR_Konvergenz} can also be observed empirically in Table \ref{tab:Expectation-of-the-MR} for normal distributed as well as non-normal distributed random variables. 

\begin{figure}[H]
\begin{centering}
\includegraphics[width=7cm]{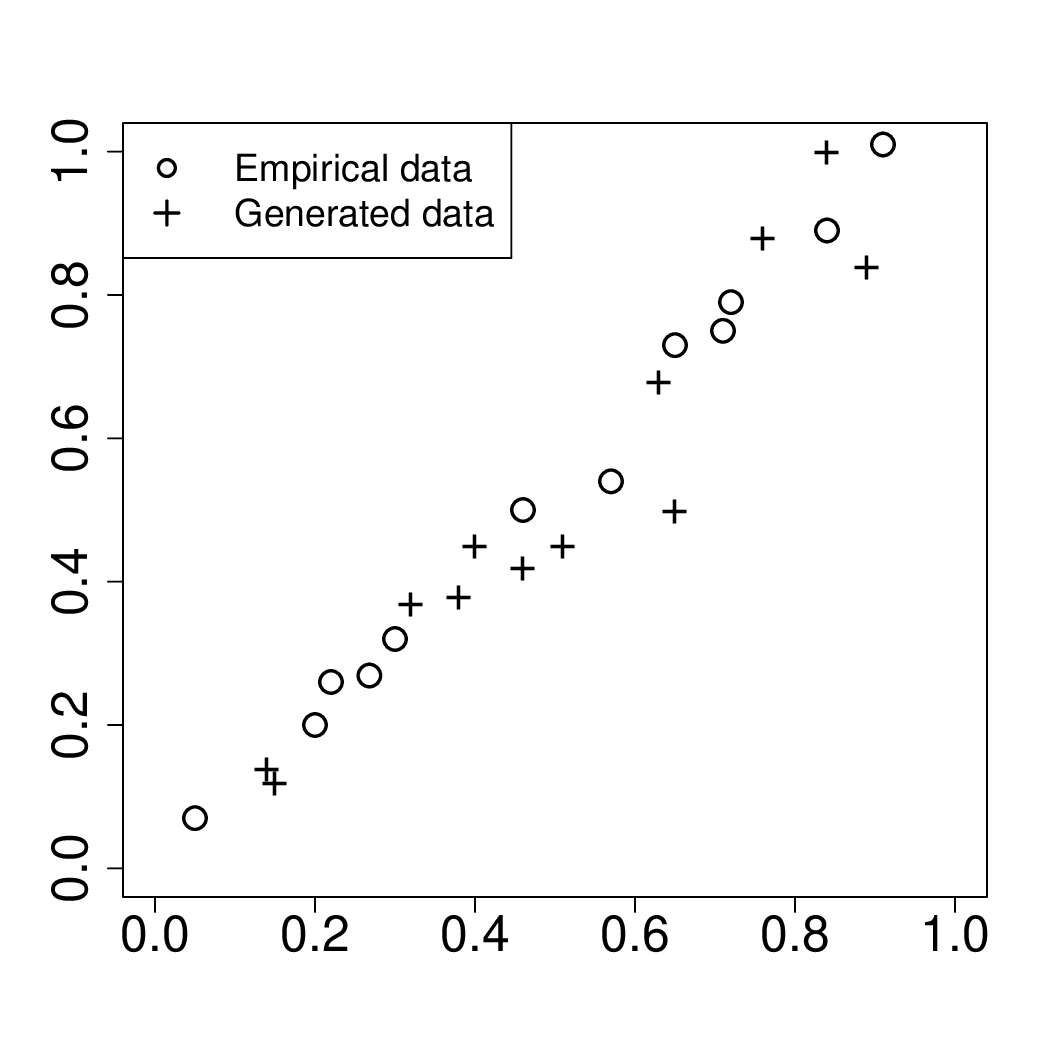}~\includegraphics[width=7cm]{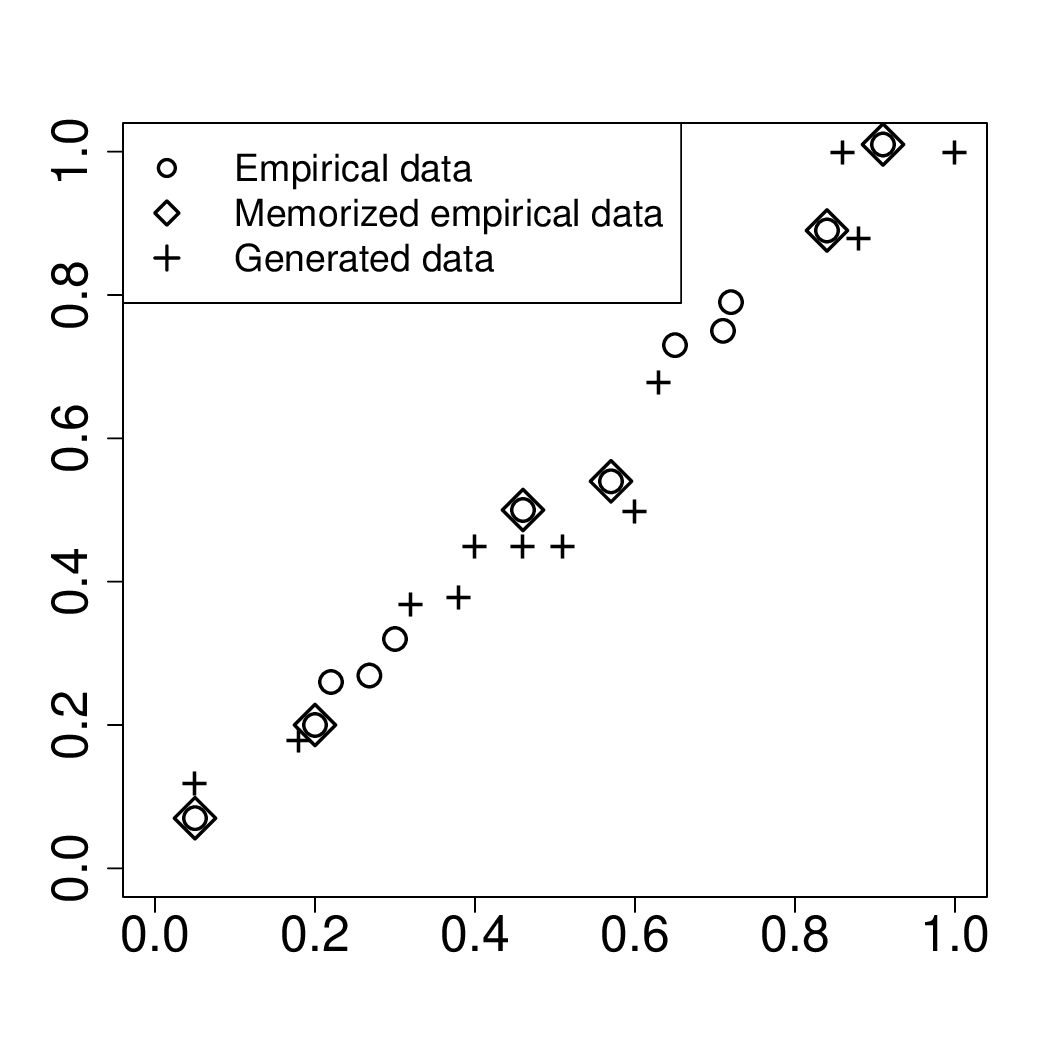}\par\end{centering}
\caption{\protect\label{fig:Example-for-the-MR}Example for the benefit of
the memorization ratio; on the left: $T_{NN1,k}=0.02$, $\Pi_{M,N}^{\rho}=0$;
on the right: $T_{NN1,k}=0.02$, $\Pi_{M,N}^{\rho}=0.5$, where $k=3$
and $\rho=0.25$.}
\end{figure}

The benefit of the memorization ratio can be visualized in Figure \ref{fig:Example-for-the-MR}. 
Some empirical data is compared to two different sets of generated data. Both generated data sets are fine according to the statistic $T_{NN1,k}$, i.e., are similarly distributed  as the empirical data. However, one generated data set leads to a high memorization ratio and can thus be ruled out.

\begin{table}[H]
\begin{centering}
\begin{tabular}{>{\centering}p{3.5cm}>{\centering}p{1cm}>{\centering}p{1.7cm}>{\centering}p{1.7cm}>{\centering}p{1.7cm}>{\centering}p{1.7cm}}
\toprule 
 & $\rho$ & Two corr. Gaussian & Exp. and Cauchy & $20$ uniform & Theoretical limit\tabularnewline
\midrule
\midrule 
$M=100$, $N=100$ & $0.25$ & 0.200 & 0.207 & 0.275 & 0.200\tabularnewline
\midrule 
$M=100$, $N=400$ & $0.25$ & 0.496 & 0.506 & 0.598 & 0.500\tabularnewline
\midrule 
$M=200$, $N=100$ & $0.25$ & 0.109 & 0.114 & 0.159 & 0.111\tabularnewline
\midrule 
$M=100$, $N=100$ & $0.5$ & 0.333 & 0.338 & 0.392 & 0.333\tabularnewline
\midrule 
$M=100$, $N=400$ & $0.5$ & 0.661 & 0.676 & 0.706 & 0.667\tabularnewline
\midrule 
$M=200$, $N=100$ & $0.5$ & 0.198 & 0.205 & 0.234 & 0.200\tabularnewline
\bottomrule
\end{tabular}
\par\end{centering}
\caption{\protect\label{tab:Expectation-of-the-MR}Memorization ratio for different
distributions: Consider four independent random variables: let $X_{1}$,
$X_{2}$ be standard normal distributed, $X_{3}$ exponentially distributed
with rate one and $X_{4}$ Cauchy distributed with location zero and
scale one. Let $E_{1},...,E_{M},G_{1},...,G_{N}$ be independent copies
of $\left(X_{1},\gamma X_{1}+\sqrt{1-\gamma^{2}}X_{2}\right)$ with
correlation $\gamma=0.75$ for the column \textquotedblleft Two corr.
Gaussian\textquotedblright . Let them be independent copies of $(X_{3},X_{4})$
for the column \textquotedblleft Exp. and Cauchy\textquotedblright .
For the column \textquotedblleft 20 uniform\textquotedblright , let
them be independent copies of $20$ independent, uniform distributed
random variables. We average the memorization ratio over $100$ simulations.
Each entry in the table comes with a standard error of at most $\pm0.006$. The values in the
table hardly change choosing the norm $\left\Vert \cdot\right\Vert $
for the memorization ratio as defined in Remark \ref{rem:=000020indis}.}
\end{table}

\begin{rem}
\label{rem:=000020indis}If a generated data point $G=(G^{1},...,G^{d})$ is identical to an empirical data point $E=(E^{1},...,E^{d})$ in all dimensions except, say, the first dimension, it depends on the application whether $E$ should be deemed 'memorized' or not. 
Choosing the Euclidean distance as in Definition \ref{def:MR}, we obtain
\[
\left\Vert G-E\right\Vert _{2}=|G^{1}-E^{1}|
\]
and $E$ would usually \emph{not} be considered as memorized if $G^{1}$ and $E^{1}$ are different enough. 
In some applications, this might not be desirable, i.e., $E$ should be deemed memorized since it is almost indistinguishable from $G$. 
One solution would be to choose another metric for the memorization ratio, e.g, based on the norm
\[
\left\Vert x\right\Vert :=\underbrace{\frac{2}{\sqrt{\pi}}\left(\frac{\Gamma\left(\frac{d}{2}+1\right)}{d!}\right)^{\frac{1}{d}}}_{=:\beta(d)}\left(\left|x^{1}\right|+...+\left|x^{d}\right|\right),\quad x\in\mathbb{R}^{d}.
\]
A unit ball under $\left\Vert \cdot\right\Vert $ has the same volume as a unit ball under $\left\Vert \cdot\right\Vert _{2}$. 
Since $\beta(d)\to0$ for $d\to\infty$, we then have
\[
\left\Vert G-E\right\Vert =\beta(d)|G^{1}-E^{1}|\to0,\quad d\to\infty
\]
and $E$ is always deemed memorized if $d$ is large enough. 
This is also confirmed numerically in Table \ref{tab:L1L2}.
\end{rem}
\begin{table}[H]
\begin{centering}
\begin{tabular}{|c|c|c|}
\hline 
$d$ & MR with $\left\Vert \cdot\right\Vert $ & MR with $\left\Vert \cdot\right\Vert _{2}$\tabularnewline
\hline 
\hline 
10 & $0.02$ & $0.0$\tabularnewline
\hline 
20 & $0.98$ & $0.002$\tabularnewline
\hline 
30 & $1.0$ & $0.02$\tabularnewline
\hline 
\end{tabular}
\par\end{centering}
\caption{\protect\label{tab:L1L2}Take $M=N=1000$, $E_{1},...,E_{M}$ are
independent and multivariate normally distributed in $\mathbb{R}^{d}$
with mean zero and covariance matrix equal to the identity. We set $G_{1}$
independent of $E_{1},...,E_{M}$ but with mean equal to $(8,...,8)$
and covariance matrix equal to the identity and $G_{i}:=E_{i}$, $i=2,...,M$.
The memorization ratio (MR) is obtained for $\rho=\frac{1}{2}$ and
using the norm $\left\Vert \cdot\right\Vert $ and $\left\Vert \cdot\right\Vert _{2}$.
The entries in the table come with a standard error of at most $\pm0.0005$.}
\end{table}

\begin{rem}
Motivated by a reviewer's remark, we comment on a potential generalization of Theorem \ref{thm:MR_Konvergenz}: 
the convergence result should also hold for identically distributed $E_m$ with a finite range dependence (i.e.\ all $E_{m+k}$ are independent of $E_m$ for $k \ge K$); thus allowing for finitely overlapping estimation windows in time series.
Figure \ref{fig:memorization-ratio-for} indicates that such a statement should hold true, but we leave a potentially more technically involved proof for future research. 
\end{rem}
\begin{rem}
Denote by $P_{E}$ the distribution of $E$ and by $P_{G}$ the distribution of $G$. 
By $P_{E}^{M}$ and $P_{G}^{N}$ we denote the empirical distribution of an empirical sample of size $M$ and a generated sample of size $N$. 
\citet{arora2017generalization} and \citet{cont2022tail} define the generalization error of $P_{G}$ by
\begin{equation}
\left|d\left(P_{E}^{M},P_{G}^{N}\right)-d\left(P_{E},P_{G}\right)\right|,\label{eq:d}
\end{equation}
where $d$ represents a divergence, e.g., the Wasserstein distance, quantile divergence, the neural net distance, etc., see \citet{arora2017generalization,cont2022tail} and references therein. 
A large probability of a small generalization error means that the generator has a similar performance with the empirical distributions and the true distributions, i.e., generalizes well from the training data set.  
However, this definition does not capture the memorization effect: Assume $P_{E}$ and $P_{G}$ are one-dimensional standard normal distributions.
Consider some empirical data of size $M$ distributed according to $P_{E}$ and generate data by simply bootstrapping (draw $M$ times independently with repetition) from the empirical data. 
If $d$ is the Wasserstein distance, the generalization error of $P_{G}$ converges to zero in probability for $M\to\infty$ but the training sample is memorized since bootstrapping does not generate new, unseen samples.
\end{rem}

\section{Numerical experiments}\label{NumerExp}

In this section, consider three different setups: in Section \ref{subsec:SP500}, we apply the concepts to one-dimensional real market data for illustrating purpose. In Section \ref{subsec:Dependent-data}, we work with four-dimensional, auto-correlated simulated data. In Section \ref{subsec:Auto-Encoders-and-interest}, we investigate in detail a purely data driven setup and deal with high-dimensional, real-market interest rate data and autoencoders.

\subsection{Illustrating one-dimensional example}\label{subsec:SP500}

In this section we illustrate in a simple setting how the memorization ratio and the test statistic $T_{NN1,k}$ can be used to evaluate various approaches to generate new samples from a training set.

We consider a training set consisting of the yearly log-returns of the S\&P 500 from 1997 to 2011 and a test set of the years 2012 to 2023. 
We discuss several ways to generate new samples from the training data, i.e., bootstrapping, a parametric approach and a ML-based method:
\begin{itemize}
\item Bootstrapping: to generate new samples, we simply draw independently with repetition from the training set (with $N = M$). The probability that a scenario from the training set shows up in the bootstrap-sample, and is hence memorized, is $1-\left(1-\frac{1}{M}\right)^{M}\approx1-e^{-1}=0.63$, cf.\ Table~\ref{tab:In-sample-and-out-of-sample}.
\item Parametric model: we use the normal distribution with mean and standard deviation equal to the empirical mean and standard deviation of the training data set to generate new samples.
\item Data-driven with kernel-smoothing: we calibrate a kernel-smoothed cumulative distribution function with Gaussian kernel with global bandwidth $h>0$ to the training data and use that distribution to generate new samples, i.e., we use the CDF
\[
F_{h}(x)=\frac{1}{M}\sum_{m=1}^{M}\Phi\left(\frac{x-E_{m}}{h}\right),\quad x\in\mathbb{R},
\]
where $\Phi$ is the CDF of a standard normal random variable and $E_{1},...,E_{M}$ is the training data.
\end{itemize}

\begin{figure}[H]
\begin{centering}
\includegraphics[scale=0.5]{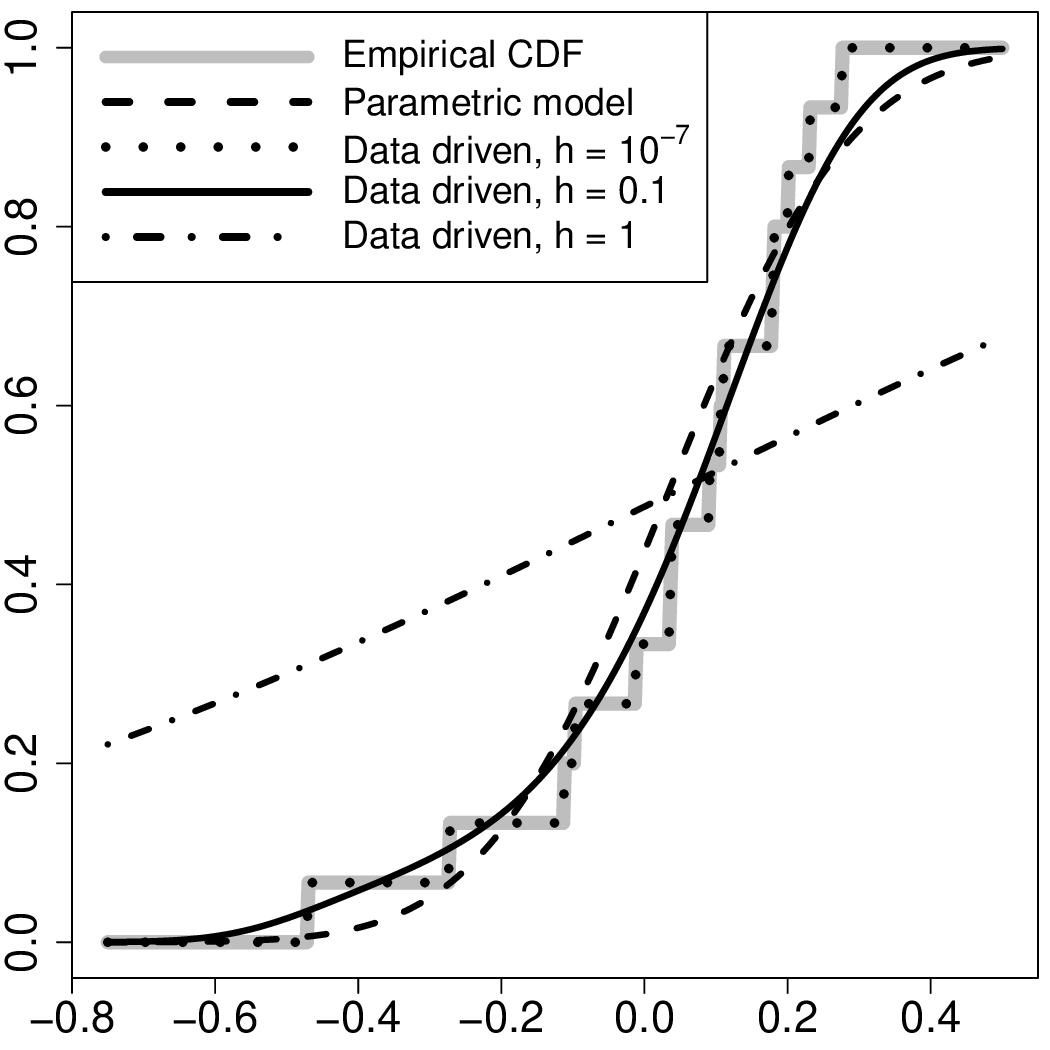}
\par\end{centering}
\caption{\protect\label{fig:Several-CDFs-to}Several CDFs to generate new data.}
\end{figure}

The empirical CDF of the training set and the CDFs of the various data generation methods are shown in Figure \ref{fig:Several-CDFs-to}. In Table \ref{tab:In-sample-and-out-of-sample}, we provide the test statistic $T_{NN1,k}$ and the memorization ratio for the generated data versus the training set (in-sample) and the generated data versus the test set (out-of-sample). 
We observe that all data generation methods except the one based on kernel-smoothing with $h=0.1$ and the parametric model show severe deficits:
Bootstrapping and kernel-smoothing with a very small $h$ result in a very high memorization ratio: overfitting takes place because the empirical CDF is just memorized. 
Underfitting occurs when the bandwidth for kernel-smoothing is too large, because the kernel-smoothed CDF does not reproduce the empirical distribution function very well. 
In general, underfitting can be easily discovered by large values of the test statistic $T_{NN1,k}$. 
We note that in this very simple setting, the parametric model fits the distribution reasonably well. 
Unfortunately, especially for larger dimensions, adequate parametric models cannot always be determined.
For example, in Section \ref{subsec:Auto-Encoders-and-interest}, we use an autoencoder-based approach to generate yield curves (of dim 31); a setup where most parametric models are known to have some issues, especially in the dependence structure of different maturities.

\begin{table}[H]
\begin{centering}
\begin{tabular}{>{\centering}p{3.5cm}>{\centering}p{1.7cm}>{\centering}p{1.5cm}>{\centering}p{2.1cm}>{\centering}p{1.5cm}}
\toprule 
Data generation Method & $T_{NN1,k}$ in-sample & MR in-sample & $T_{NN1,k}$ out-of-sample & MR out-of-sample\tabularnewline
\midrule
\midrule 
Bootstrap & 0.06 & \textbf{0.64} & 0.08 & 0.11\tabularnewline
\midrule 
Data-driven, $h=10^{-7}$ & 0.05 & \textbf{0.65} & 0.07 & 0.14\tabularnewline
\midrule 
Data-driven, $h=1$ & \textbf{0.22} & 0.08 & \textbf{0.27} & 0.07\tabularnewline
\midrule 
Data-driven, $h=0.1$ & 0.06 & 0.19 & 0.07 & 0.19\tabularnewline
\midrule 
Parametric model & 0.06 & 0.17 & 0.07 & 0.21\tabularnewline
\bottomrule
\end{tabular}
\par\end{centering}
\caption{\protect\label{tab:In-sample-and-out-of-sample}In-sample and out-of-sample
validation using the memorization ratio (MR) with $\rho=0.25$ and the
test statistic $T_{NN1,k}$ with $k=3$. In-sample ($M=N=15$): empirical data is the training set. Out-of-sample ($M=N=12$): empirical data is the test set. Unacceptable
values are displayed in bold. We average over $100$ simulations.
Since $M=N$, under \ref{H0}, the memorization ratio
converges to $\frac{\rho}{\rho+1}=0.2$ for in-sample validation and out-of-sample validation.
The entries in the table come with a standard error of at most $\pm0.01$.}
\end{table}

\begin{figure}[H]
\begin{centering}
\includegraphics[scale=0.5]{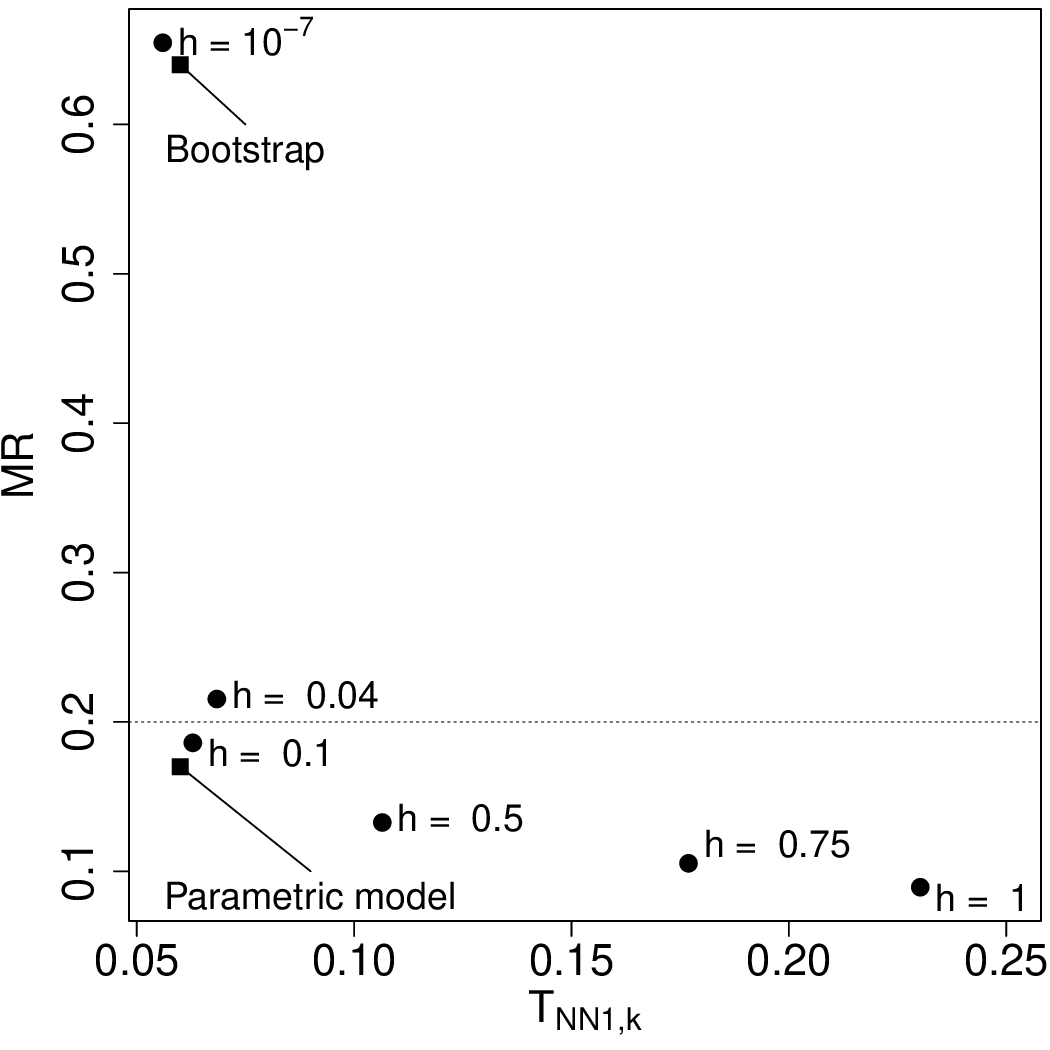}
\par\end{centering}
\caption{\protect\label{fig:MR_over_NNC}Memorization ratio (MR) with $\rho=0.25$ versus $T_{NN1,k}$ with $k=3$. New samples are generated by a data-driven approach using kernel-smoothing with different bandwidths $h>0$, bootstrapping and a Monte Carlo simulation via a parametric model. Generated scenarios are compared to the training set. The theoretical limit of the memorization ratio is $0.2$.  We average over $100$ simulations.}
\end{figure}

Figure \ref{fig:MR_over_NNC} shows the memorization ratio versus the statistic $T_{NN1,k}$, where the empirical data is the training set and new samples are generated by a data-driven approach using kernel-smoothing with different bandwidths $h>0$, bootstrapping and a Monte Carlo simulation via a parametric model. The optimal scenario generator can be chosen such that the memorization ratio is close to its theoretical limit and the statistic $T_{NN1,k}$ is small.

\subsection{Dependent data}\label{subsec:Dependent-data}

\begin{figure}[H]
\begin{centering}
\includegraphics[scale=0.5]{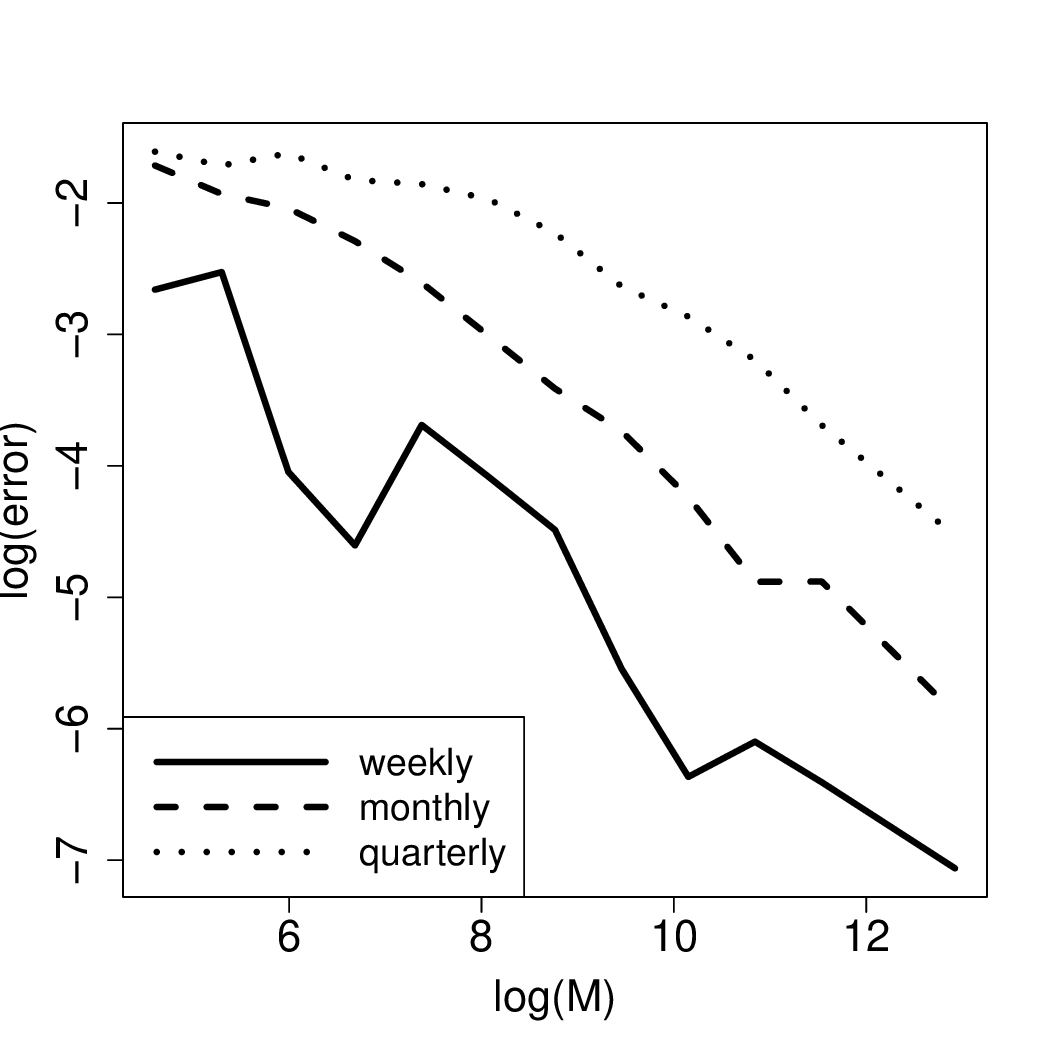}
\par\end{centering}
\caption{\protect\label{fig:memorization-ratio-for}Memorization ratio for three different
window sizes for increasing $M$ and $N$. We choose $N=M$ and $\rho=0.25$ and consider
a correlated geometric Brownian motion in four dimensions as data
generation process. The error is the absolute difference
of the memorization ratio and its limit.}
\end{figure}

For illustration purposes, we simulate four dimensional financial daily data by a (correlated) geometric
Brownian motion. 
Technical details can be found in Appendix \ref{sec:Technical-details:-dependent}.
We apply a rolling (estimation) window to this data in order to construct financial scenarios over a larger time horizon.
Let $w\in\mathbb{N}$ be some appropriate window size, e.g., $w=5,$ $w=21$ and $w=63$ corresponds to weekly, monthly or quarterly data. 
If $w>1$, the data becomes auto-correlated in the time-dimension, due to window overlapping.
We see in Figure \ref{fig:memorization-ratio-for} that the memorization ratio still seems to converge (even though the independence assumption in \ref{H0} is violated) with the same order of convergence.

\subsection{Interest rate scenario generation by autoencoders}\label{subsec:Auto-Encoders-and-interest}

After illustrating the main principles based on low-dimensional and simple examples, we now consider a purely data-driven setup.
We start with interest rate data on a daily basis from 2000-08-01 to 2023-01-31 (more specifically German government rates, taken from the website of the Deutsche Bundesbank\footnote{\small{\url{https://www.bundesbank.de/dynamic/action/de/statistiken/zeitreihen-datenbanken/zeitreihen-datenbank/759778/759778?listId=www_skms_it03a}}\normalsize\,
}), containing 5716 data points of dimension 31 (maturities 6M, 1Y, 2Y, \dots, 30Y). 
Since we are interested in yield curve scenarios over the period of a month, we take the observations 1, 21, 41, \dots 5701, leading to 284 independent 1M-differences (one-month differences).
Given generated 1M-differences, annual interest rate scenarios for SCR calculations can then be obtained by concatenating twelve 1M-differences.

We use the first 260 difference curves as training data and hold out the remaining 24 difference curves as validation data.
As autoencoder setup, we select a fully-connected feed-forward autoencoder with three hidden layers of size 62, 2, and 62 (note that input and output size is 31). 
Since the number of neurons as well as the number of latent factors (in our setup the dimension of the latent factors $Z$ is two) do not significantly change the results, we have opted for this setup, especially for two latent factors, for simplicity of illustration.
Of course, allowing for more latent factors further decreases the reconstruction error.

\begin{figure}[H]
\begin{centering}
\includegraphics[scale=1.3]{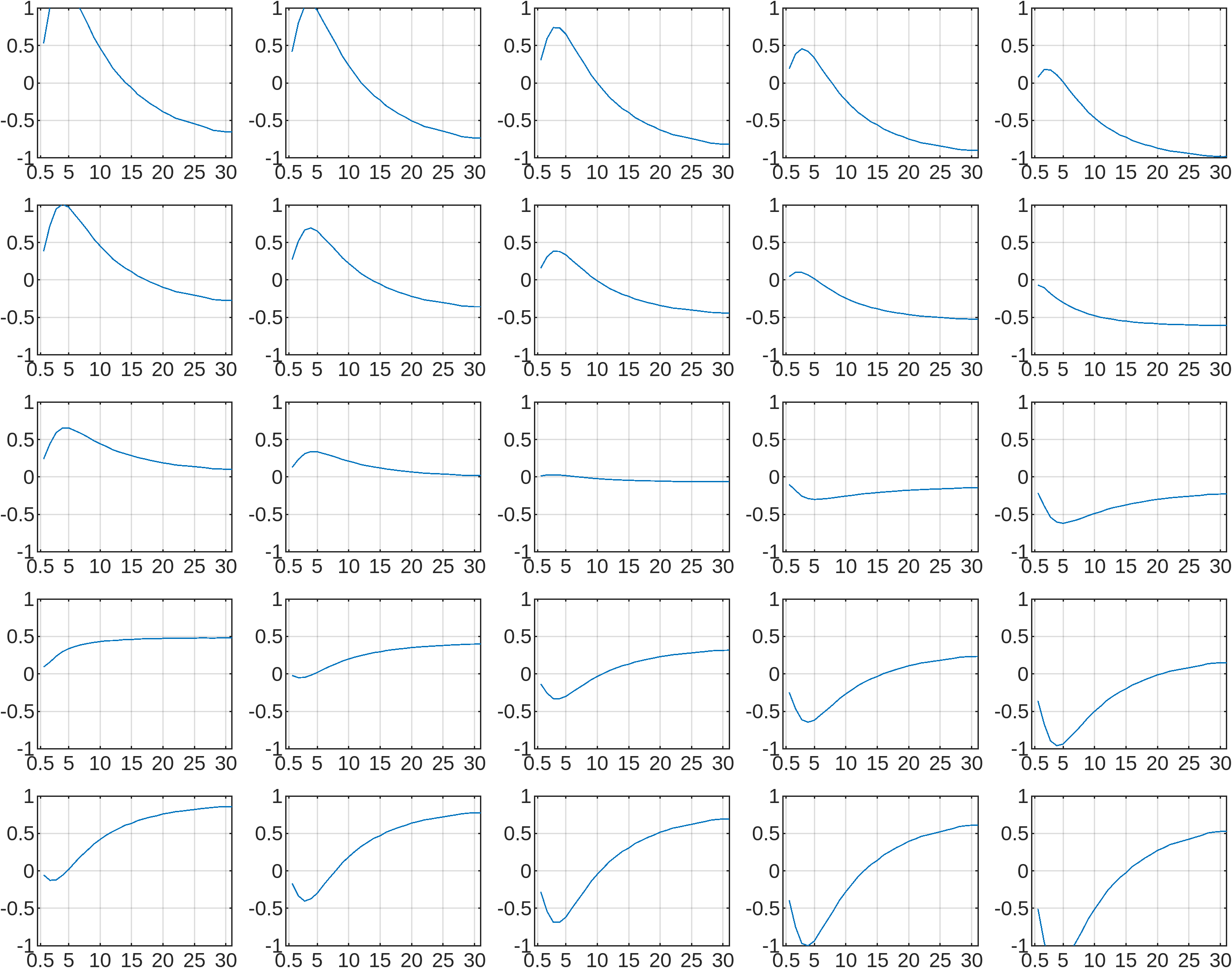}
\par\end{centering}
\caption{Different types of yield curves (1M-differences) obtained for $Z \in \{-2,-1,0,1,2\}^2$.}\label{fig:illustrationAE}
\end{figure}

For the training of the autoencoder, the L-BFGS algorithm has been applied to minimize the mean-absolute-error. 
All implementations in this section have been carried out in Matlab R2024b, mainly based on the Deep Learning Toolbox. 
With two latent factors, the final in-sample reconstruction loss is approx.\ 2 basis points (bp), and out-of-sample 2bp as well, which shows that encoding/decoding can be performed with quite good quality.
With four latent factors, it can be further decreased (in-sample and out-of-sample) to 1bp.
Figure~\ref{fig:illustrationAE} displays the different type of yield curves the autoencoder can generate when varying the latent factors.

\begin{figure}[H]
\begin{centering}
\includegraphics[scale=0.90]{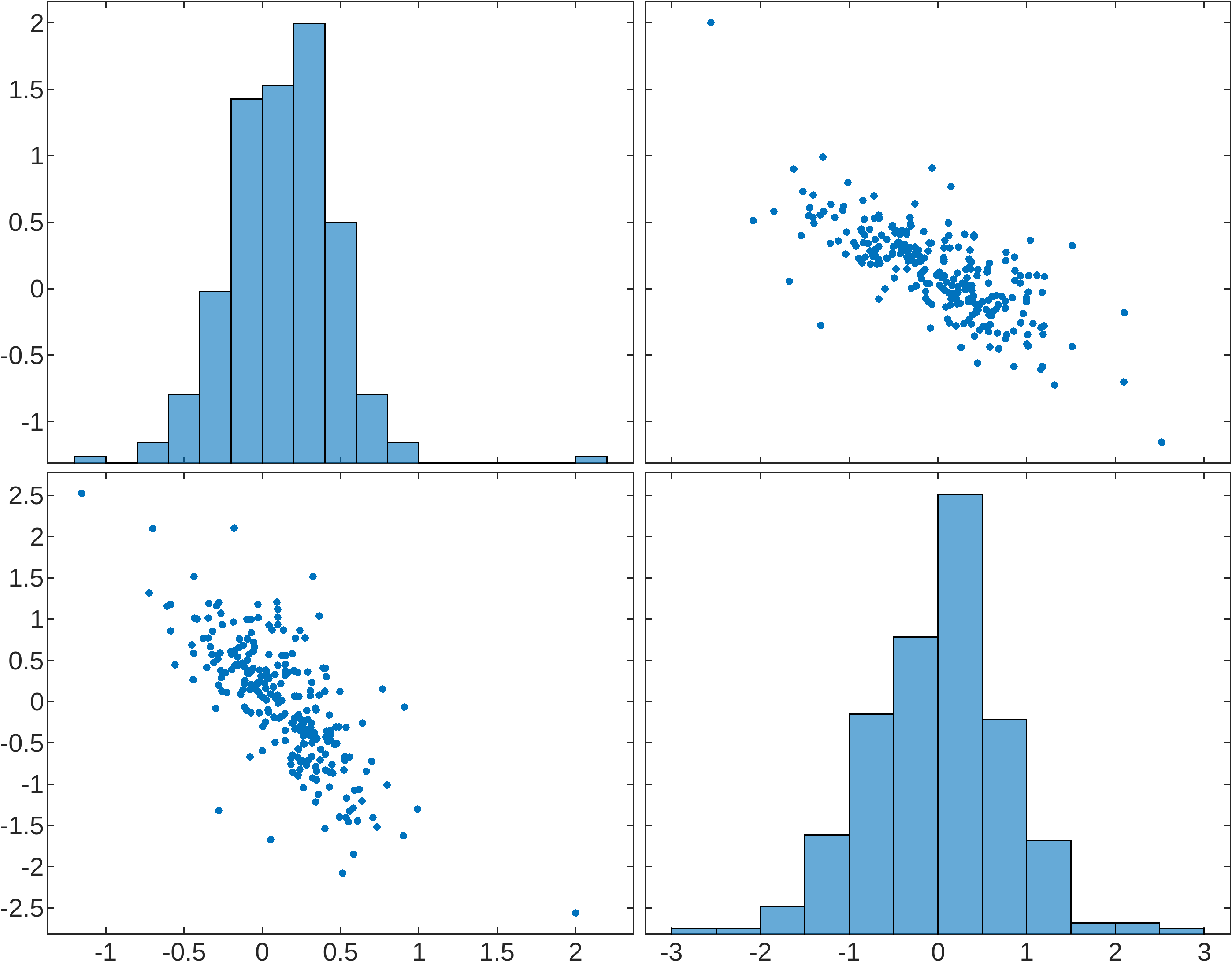}\\
\,\\
\includegraphics[scale=0.90]{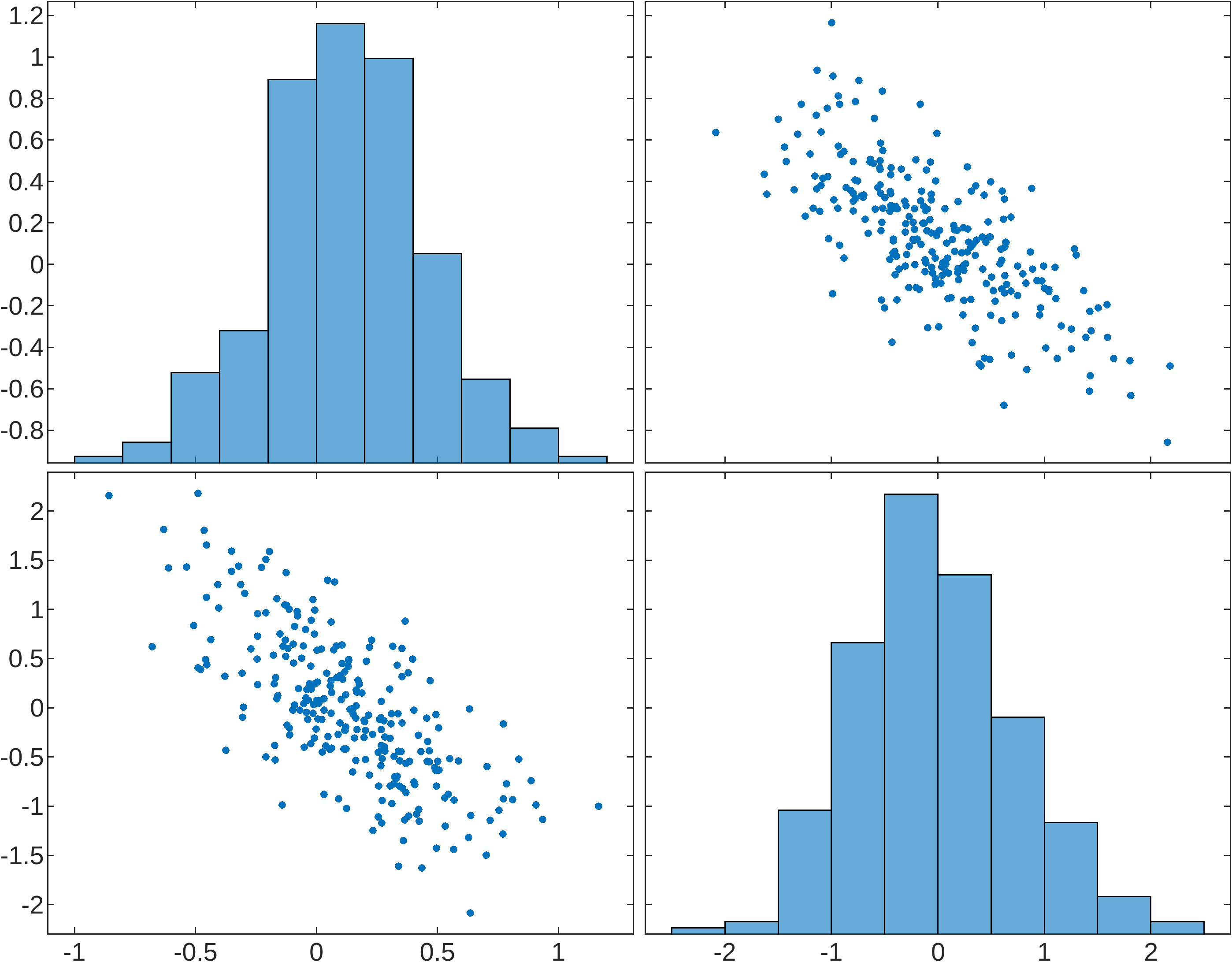}
\par\end{centering}
\caption{Top panel: marginal distribution and scatter plot of the 260 observations of the latent factors corresponding to the 260 yield curve differences. Bottom panel: marginal distribution and scatter plot of 260 randomly generated scenarios of the latent factors.}\label{fig:AE-F1-latent-vars}
\end{figure}

In Figure~\ref{fig:AE-F1-latent-vars}, top panel, we have illustrated the 260 latent factors corresponding to the 260 1M-difference curves. 
In the bottom panel of Figure~\ref{fig:AE-F1-latent-vars}, we have illustrated a random sample for $Z$ according to a 2-dim normal distribution with the same mean and covariance matrix as the latent factors of the training data, whose distribution remains unknown.

Based on the trained autoencoder and its split into an encoder and decoder part, we then generate the following data sets:
\begin{itemize}
\item $S_1$: In this set, we collect the 260 original 1M-difference curves.
\item $S_2$: In this set, we collect the 260 encoded and afterwards decoded 1M-difference curves.
\item $S_3$: This represents the main ML-based generated data, generated by replacing the 260 realizations of the 2-dim latent factor by 260 random realizations of a 2-dim normal distribution with the same mean and covariance matrix as the latent factors.
\item $S_4$, $S_5$ and $S_6$: New curves are obtained by distorting the elements of $S_1$ by adding some independent noise with increasing noise level.
\end{itemize}
Here, the noise for the set $S_5$ comes from a 31-dim normal distribution with the same mean and the same covariance matrix as the 31-dim reconstruction error of the autoencoder, while the noise for set $S_6$ is multiplied by 10 (in terms of variance), whereas it is divided by 10 for $S_4$.
\begin{figure}[H]
\begin{centering}
\includegraphics[scale=0.5]{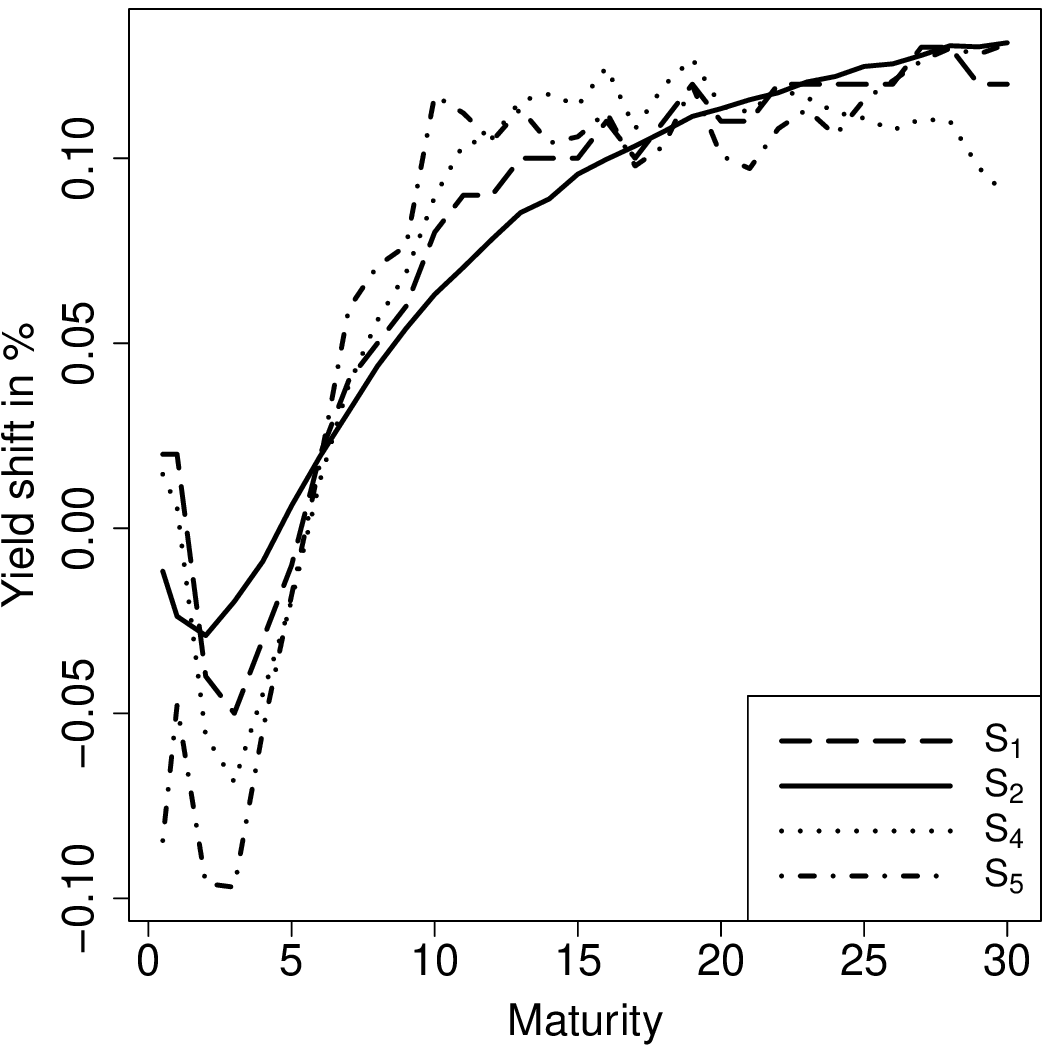}
\par\end{centering}
\caption{The dashed line represents $E_{20}$, i.e.,\ the 20-th curve from the training data ($S_1$), while the solid line represents its encoded-decoded counterpart $Y_{20}$, i.e.,\ the 20-th element in $S_2$. The other two curves are distorted versions of $E_{20}$, with low noise ($S_4$) and with a noise level similar to $Y_{20}$ ($S_5$).}\label{fig:curves}
\end{figure}
We have illustrated the 1M-difference curves in the sets $S_i$ in Figure~\ref{fig:curves} by a specific instance (instance no.\ 20), which represents a typical situation.
We can immediately observe that the corresponding curve from $S_4$ is very close to its original counterpart in $S_1$, since the noise is very small. 
The corresponding curve from $S_5$ is already further away due to the higher noise, especially at the short end of the yield curve.
Note that we did not visualize curves from $S_6$ because they are too far off.
While the encoded-decoded curves from $S_2$ are also close to the original counterpart (by construction), it can still be observed that curves from $S_2$ seem to be smoother than all other curves -- an often observed feature of autoencoders.
Let us recall that the difference between the curves in $S_1$ and $S_2$ is solely due to the reconstruction error of the autoencoder, which is on average 1.9bps per tenor, and hence sufficiently small.
$S_3$ has also been left out of the visualization, as this would not make sense, since the randomly generated data is not in a 1-to-1 correspondence to data in the training set.

In Table~\ref{tab:AE-results} we have summarized the results of the quantitative comparison of $S_2$, $S_3$, \dots, $S_6$ with $S_1$:
As expected, $S_2$ is not too far from $S_1$ in terms of $T_{NN1,k}$, while the memorization ratio is  noticeably higher than its limit 4/9 under \ref{H0}, which is a clear indicator for memorization.
Due to the reconstruction error, the memorization error remains significantly beneath 1, as well as $T_{NN1,k}$ remains significantly above 0.
We can further observe ($T_{NN1,k} = 0.15$) that the generated data $S_3$ is at least as close to the empirical data as is the encoded-decoded data $S_2$, while having a much better memorization ratio.
Concerning the distorted data, $S_5$ has a slightly better $T_{NN1,k}$ than $S_2$ and $S_3$, while $S_4$ is the closest to $S_1$, and $S_6$ is furthest away from $S_1$ due to its higher level of noise.
The memorization ratio decreases with increasing level of noise, which also meets expectations. 
In summary, we can observe a good performance of the autoencoder-based scenario generation in the column $S_3$: we get a similar performance as by disturbing the original data with the same level of noise (i.e.\ $S_5)$ -- but now, more meaningful (i.e.\ smoother) yield curves are constructed and the dimension for data generation has been significantly reduced (from 31 to 2).
%
\begin{table}[H]
\begin{centering}
\begin{tabular}{>{\centering}p{3.5cm}>{\centering}p{1.7cm}>{\centering}p{1.7cm}>{\centering}p{1.7cm}>{\centering}p{1.7cm}>{\centering}p{1.7cm}}
\toprule 
$S_1$ compared to & $S_2$ & $S_3$ & $S_4$ & $S_5$ &  $S_6$\tabularnewline
\midrule
\midrule  
$T_{NN1,k}$, $k=5$ & 0.17 & 0.15 & 0.09 & 0.11 & 0.26 \tabularnewline
\midrule
\midrule 
MR, $\rho = 0.80$ & 0.63 & 0.51 & 0.93 & 0.49 & 0.09 \tabularnewline
\bottomrule
\end{tabular}
\par\end{centering}
\caption{\protect\label{tab:AE-results}In-sample comparison of the memorization ratio (MR) and the test statistic $T_{NN1,k}$ ($N=M=260$). 
Note that under \ref{H0} the memorization ratio (MR) converges to $\frac{\rho}{\rho+1}$, i.e., to $4/9$.
We average over $100$ simulations and entries in the table come with a standard error of at most $\pm0.01$.}
\end{table}
To validate our results, we run the following out-of-sample test: Let $S_{1}^{\ast}$ denote
the validation set consisting of 24 1M-difference curves. Let $S_{3}^{\ast}$
be ML-based generated data (similar to $S_{3}$), generated by replacing
the 2-dim latent factor of the autoencoder by 24 random realizations of a corresponding 2-dim normal
distribution. The $T_{NN1,k}$ statistic with $k=5$ between $S_{1}^{\ast}$ and
$S_{3}^{\ast}$ is 0.04, i.e., also out-of-sample, we accept the hypothesis
that data generated by the autoencoder has the same distribution as
the underlying distribution of the empirical data. 

Except memorization caused by randomness, it is impossible for the
generated set $S_{3}^{\ast}$ to memorize any element of $S_{1}^{\ast}$
since the autoencoder has never seen the validation set $S_{1}^{\ast}$
during training. The memorization ratio with $\rho=0.8$ between $S_{1}^{\ast}$ and
$S_{3}^{\ast}$ is 0.5, which is almost the same as the memorization
ratio between $S_{1}$ and $S_{3}$. The memorization ratio is only slightly
above the theoretical limit of $4/9$ both in-sample and
out-of-sample. From this out-of-sample validation, we conclude that
the autoencoder does not systematically memorize any training data,
since the memorization ratios for in-sample and out-of-sample data
are almost equal and close to its theoretical limit. 
%
%
%

\begin{figure}[H]
\begin{centering}
\includegraphics[scale=1]{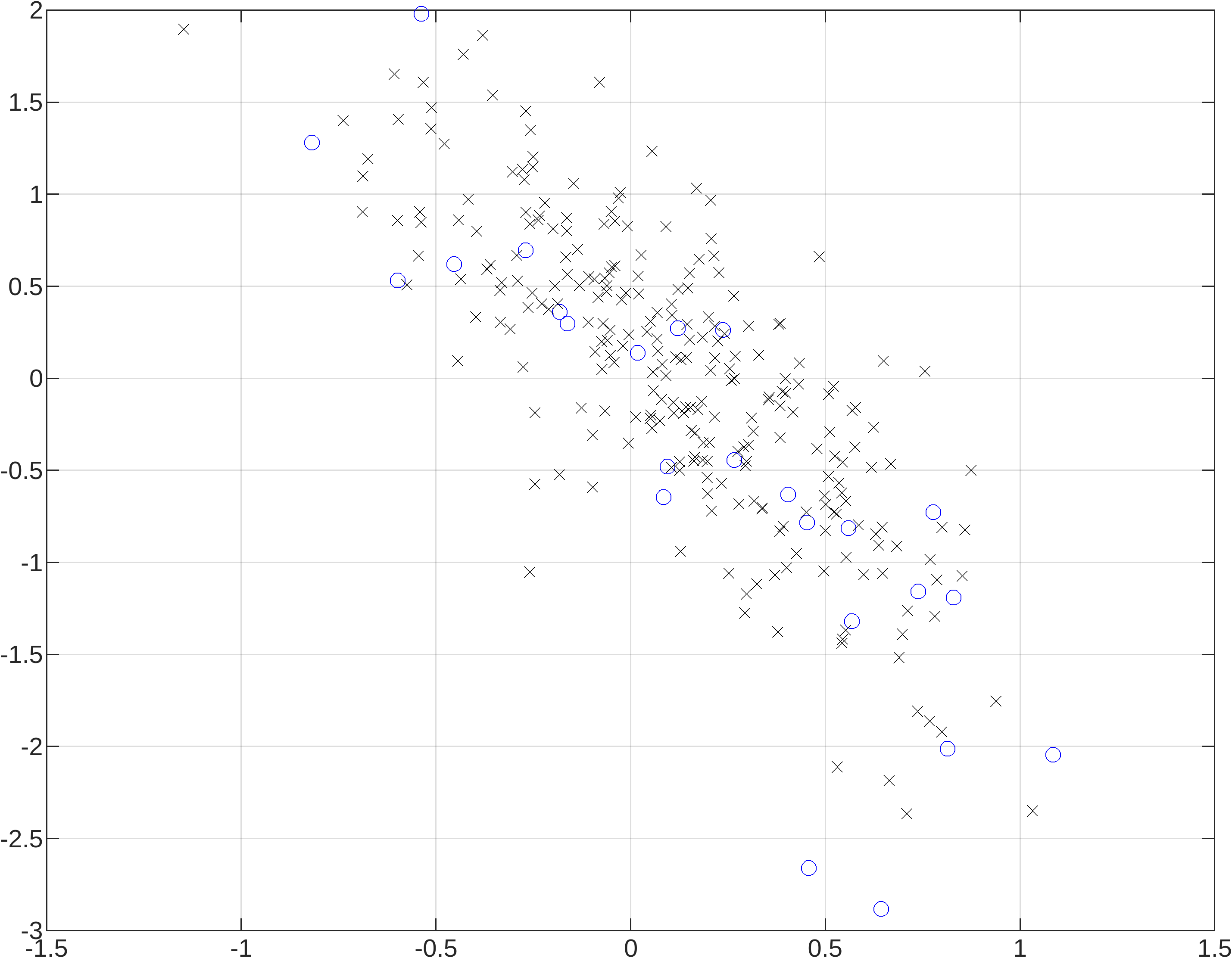}
\par\end{centering}
\caption{The black 'x' represent data in $S_3$, while the blue 'o' represent data in $S_1^*$.
Note that the figure does not show the original 31-dim data, but its encoded 2-dim representation.}\label{fig:oosCoverage}
\end{figure}

Regarding the out-of-sample performance of the scenario generator and its adequacy for risk calculations, we need to prove that the generated scenarios $S_3$ match the out-of-sample distribution of $S_1^*$ sufficiently well: 
The $T_{NN1,k}$ statistic with $k=5$ between the out-of-sample data $S_{1}^{\ast}$ and the generated scenarios $S_{3}$ is on average 0.04 -- indicating that the generated distribution predicts the out-of-sample distribution quite well.
A visual inspection of the data in Figure~\ref{fig:oosCoverage} shows that the out-of-sample scenarios are indeed all (perhaps except for two out-of-sample scenarios) covered by the generated scenarios.
Note that 31-dim data cannot be displayed in a 2-dim plot without some projection. 
We have used coded values, encoded by the autoencoder, but we could have used classical multidimensional scaling for illustration purposes as well.
Last but not least, the memorization ratio between $S_1^*$ and $S_3$ (for $\rho = 0.8$) is 0.88 and thus very close to the theoretical limit of 0.90. As desired, the generated data $S_3$ almost completely covers the validation data $S_1^*$.
The memorization ratio and the $T_{NN1,k}$ statistic together imply that the generated data does not contain any point of $S_1^*$ (except by chance), but produces a distribution very close to the out-of-sample distribution.

\section{Conclusions}\label{conc}

When using a ML-based model
for regulatory purposes, such as modeling of risks under Solvency 2,
additional validation techniques need to be applied to take into account
the ML nature of these techniques. For validation purposes, we have provided
two specialized measures that can be used to evaluate the performance
of the scenario generation: nearest neighbor coincidences (measuring
the alignment of the multivariate distributions of the risk factors, see Section \ref{subsec:NNC}) and a
new measure, the memorization ratio (measuring overfitting, see Section \ref{subsec:MR}). We show
that the memorization ratio converges
when the amount of data increases.  Numerical experiments on simulated and real market data using several data generation methods (bootstrapping, kernel-smoothing, Monte Carlo and autoencoders) show that the memorization ratio is able to detect the memorization effect.

\subsection*{Acknowledgments and funding}

We thank two referees for many helpful comments that significantly improved the paper. Solveig Flaig would like to thank Deutsche Rückversicherung AG for the funding of this research. 
Opinions, errors and omissions are solely those of the authors and do not represent those of Deutsche Rückversicherung AG or its affiliates.

\subsection*{Disclosure statement. }

The authors report there are no competing interests to declare. 

\subsection*{Availability of data and materials}

All real data analyzed during this study are publicly available. 
URLs are included in this article.

\subsection*{Code availability}

The full code of the experiments presented in this paper is available for review by contacting the corresponding author.

\bibliographystyle{plainnat}
\phantomsection\addcontentsline{toc}{section}{\refname}\bibliography{biblio}

\begin{thebibliography}{55}
\providecommand{\natexlab}[1]{#1}
\providecommand{\url}[1]{\texttt{#1}}
\expandafter\ifx\csname urlstyle\endcsname\relax
  \providecommand{\doi}[1]{doi: #1}\else
  \providecommand{\doi}{doi: \begingroup \urlstyle{rm}\Url}\fi

\bibitem[Adesi(2014)]{adesi2014simulating}
Giovanni~Barone Adesi.
\newblock \emph{{Simulating security returns: A filtered historical simulation
  approach}}.
\newblock Springer, 2014.

\bibitem[Albeanu et~al.(2008)Albeanu, Ghica, and
  Popentiu-Vladicescu]{albeanu2008using}
Grigore Albeanu, Manuela Ghica, and Florin Popentiu-Vladicescu.
\newblock {On using bootstrap scenario-generation for multi-period stochastic
  programming applications}.
\newblock \emph{Int. J. Comput. Commun. Control}, 3:\penalty0 156--161, 2008.

\bibitem[Arian et~al.(2022)Arian, Moghimi, Tabatabaei, and
  Zamani]{arian2022encoded}
Hamid Arian, Mehrdad Moghimi, Ehsan Tabatabaei, and Shiva Zamani.
\newblock {Encoded Value-at-Risk: A machine learning approach for portfolio
  risk measurement}.
\newblock \emph{Mathematics and Computers in Simulation}, 202:\penalty0
  500--525, 2022.

\bibitem[Arora et~al.(2017)Arora, Ge, Liang, Ma, and
  Zhang]{arora2017generalization}
Sanjeev Arora, Rong Ge, Yingyu Liang, Tengyu Ma, and Yi~Zhang.
\newblock {Generalization and equilibrium in generative adversarial nets
  (GANs)}.
\newblock In \emph{International conference on machine learning}, pages
  224--232. PMLR, 2017.

\bibitem[Arora et~al.(2018)Arora, Risteski, and Zhang]{arora2017gans}
Sanjeev Arora, Andrej Risteski, and Yi~Zhang.
\newblock {Do GANs learn the distribution? Some theory and empirics}.
\newblock In \emph{International Conference on Learning Representations}, 2018.

\bibitem[BaFin(2021)]{bafin2021ml}
BaFin.
\newblock {Machine learning in risk models: Characteristics and supervisory
  priorities}, 2021.
\newblock URL
  \url{https://www.bundesbank.de/resource/blob/793670/61532e24c3298d8b24d4d15a34f503a8/mL/2021-07-15-ml-konsultationspapier-data.pdf}.
\newblock [accessed on 2023/10/20].

\bibitem[Bai et~al.(2021)Bai, Lin, Raffel, and Kan]{bai2021training}
Ching-Yuan Bai, Hsuan-Tien Lin, Colin Raffel, and Wendy Chi-wen Kan.
\newblock {On training sample memorization: Lessons from Benchmarking
  generative modeling with a large-scale competition}.
\newblock In \emph{Proceedings of the 27th ACM SIGKDD Conference on Knowledge
  Discovery \& Data Mining}, pages 2534--2542, 2021.

\bibitem[Beder(1995)]{beder1995var}
Tanya~Styblo Beder.
\newblock {VaR: Seductive but dangerous}.
\newblock \emph{Financial Analysts Journal}, 51\penalty0 (5):\penalty0 12--24,
  1995.

\bibitem[B{\'e}gin(2021)]{begin2021complex}
Jean-Fran{\c{c}}ois B{\'e}gin.
\newblock {On Complex Economic Scenario Generators: Is Less More?}
\newblock \emph{ASTIN Bulletin: The Journal of the IAA}, 51\penalty0
  (3):\penalty0 779--812, 2021.

\bibitem[Bickel and Breiman(1983)]{bickel1983sums}
Peter~J Bickel and Leo Breiman.
\newblock {Sums of functions of nearest neighbor distances, moment bounds,
  limit theorems and a goodness of fit test}.
\newblock \emph{The Annals of Probability}, pages 185--214, 1983.

\bibitem[Borji(2019)]{borji2019pros}
Ali Borji.
\newblock {Pros and cons of GAN evaluation measures}.
\newblock \emph{Computer Vision and Image Understanding}, 179:\penalty0 41--65,
  2019.

\bibitem[Borji(2022)]{borji2022pros}
Ali Borji.
\newblock {Pros and cons of GAN evaluation measures: New developments}.
\newblock \emph{Computer Vision and Image Understanding}, 215:\penalty0 103329,
  2022.

\bibitem[Buch et~al.(2023)Buch, Grimm, Korn, and Richert]{buch2023estimating}
Robert Buch, Stefanie Grimm, Ralf Korn, and Ivo Richert.
\newblock {Estimating the value-at-risk by Temporal VAE}.
\newblock \emph{Risks}, 11\penalty0 (5):\penalty0 79, 2023.

\bibitem[Cadoni(2014)]{cadoni2014internal}
Paolo Cadoni.
\newblock \emph{{Internal models and Solvency II}}.
\newblock Risk Books, London, 2014.

\bibitem[Chen et~al.(2018)Chen, Li, and Zhang]{chen2018bayesian}
Yize Chen, Pan Li, and Baosen Zhang.
\newblock {Bayesian renewables scenario generation via deep generative
  networks}.
\newblock In \emph{2018 52nd Annual Conference on Information Sciences and
  Systems (CISS)}, pages 1--6. IEEE, 2018.

\bibitem[Cont et~al.(2022)Cont, Cucuringu, Xu, and Zhang]{cont2022tail}
Rama Cont, Mihai Cucuringu, Renyuan Xu, and Chao Zhang.
\newblock {Tail-GAN: Learning to Simulate Tail Risk Scenarios}.
\newblock \emph{arXiv preprint arXiv:2203.01664}, 2022.

\bibitem[Demirel and Willemain(2002)]{demirel2002generation}
Omer~F Demirel and Thomas~R Willemain.
\newblock {Generation of simulation input scenarios using bootstrap methods}.
\newblock \emph{Journal of the Operational Research Society}, 53\penalty0
  (1):\penalty0 69--78, 2002.

\bibitem[Dupont et~al.(2020)Dupont, Fliche, and Yang]{dupont2020governance}
Laurent Dupont, Olivier Fliche, and Su~Yang.
\newblock {Governance of artificial intelligence in finance}, 2020.
\newblock URL
  \url{https://acpr.banque-france.fr/sites/default/files/medias/documents/20200612_ai_governance_finance.pdf}.
\newblock [accessed on 2023/10/20].

\bibitem[Ebner et~al.(2018)Ebner, Henze, and Yukich]{ebner2018multivariate}
Bruno Ebner, Norbert Henze, and Joseph~E Yukich.
\newblock {Multivariate goodness-of-fit on flat and curved spaces via nearest
  neighbor distances}.
\newblock \emph{Journal of Multivariate Analysis}, 165:\penalty0 231--242,
  2018.

\bibitem[{European Union}(2009)]{union2009directive}
{European Union}.
\newblock {Directive 2009/136/EC of the European parliament and of the
  council}.
\newblock \emph{Official Journal of the European Union}, 337:\penalty0 11,
  2009.

\bibitem[Fernandez-Arjona(2021)]{fernandez2021neural}
Lucio Fernandez-Arjona.
\newblock {A neural network model for solvency calculations in life insurance}.
\newblock \emph{Annals of Actuarial Science}, 15\penalty0 (2):\penalty0
  259--275, 2021.

\bibitem[Fiechtner(2019)]{fiechtner2019risk}
Lucas~Benedikt Fiechtner.
\newblock {Risk Management with Generative Adversarial Networks}, 2019.
\newblock URL
  \url{https://www.researchgate.net/publication/347440941_Risk_Management_with_Generative_Adversarial_Networks}.
\newblock [accessed on 2023/10/20].

\bibitem[Flaig and Junike(2022)]{flaig2022scenario}
Solveig Flaig and Gero Junike.
\newblock {Scenario generation for market risk models using generative neural
  networks}.
\newblock \emph{Risks}, 10\penalty0 (11):\penalty0 199, 2022.

\bibitem[Gatzert and Martin(2012)]{gatzert2012quantifying}
Nadine Gatzert and Michael Martin.
\newblock {Quantifying credit and market risk under Solvency II: Standard
  approach versus internal model}.
\newblock \emph{Insurance: Mathematics and Economics}, 51\penalty0
  (3):\penalty0 649--666, 2012.

\bibitem[Goodfellow et~al.(2016)Goodfellow, Bengio, and
  Courville]{GoodBengCour16}
Ian~J. Goodfellow, Yoshua Bengio, and Aaron Courville.
\newblock \emph{{Deep Learning}}.
\newblock MIT Press, Cambridge, MA, USA, 2016.
\newblock \url{http://www.deeplearningbook.org}.

\bibitem[Gu et~al.(2021)Gu, Kelly, and Xiu]{gu2021autoencoder}
Shihao Gu, Bryan Kelly, and Dacheng Xiu.
\newblock {Autoencoder asset pricing models}.
\newblock \emph{Journal of Econometrics}, 222\penalty0 (1):\penalty0 429--450,
  2021.

\bibitem[Hendricks(1996)]{hendricks1996evaluation}
Darryll Hendricks.
\newblock {Evaluation of value-at-risk models using historical data}.
\newblock \emph{Economic policy review}, 2\penalty0 (1), 1996.

\bibitem[Henze(1988)]{henze1988multivariate}
Norbert Henze.
\newblock {A multivariate two-sample test based on the number of nearest
  neighbor type coincidences}.
\newblock \emph{The Annals of Statistics}, 16\penalty0 (2):\penalty0 772--783,
  1988.

\bibitem[Jung et~al.(2019)Jung, Mueller, Pedemonte, Plances, and
  Thew]{jung2019machine}
Carsten Jung, Henrike Mueller, Simone Pedemonte, Simone Plances, and Oliver
  Thew.
\newblock {Machine learning in UK financial services}.
\newblock \emph{Bank of England and Financial Conduct Authority}, 2019.
\newblock URL
  \url{https://www.bankofengland.co.uk/-/media/boe/files/report/2019/machine-learning-in-uk-financial-services.pdf}.
\newblock [accessed on 2023/10/20].

\bibitem[Kingma and Welling(2019)]{KingmaWelling2019}
Diederik~P. Kingma and Max Welling.
\newblock \emph{{An Introduction to Variational Autoencoders}}.
\newblock 2019.

\bibitem[Kondratyev and Schwarz(2019)]{kondratyev2019market}
Alexei Kondratyev and Christian Schwarz.
\newblock {The market generator}.
\newblock \emph{Available at SSRN 3384948}, 2019.

\bibitem[Lopez-Paz and Oquab(2016)]{lopez2016revisiting}
David Lopez-Paz and Maxime Oquab.
\newblock {Revisiting classifier two-sample tests}.
\newblock \emph{arXiv preprint arXiv:1610.06545}, 2016.

\bibitem[Makhzani et~al.(2015)Makhzani, Shlens, Jaitly, and
  Goodfellow]{MakhzaniSJG15}
Alireza Makhzani, Jonathon Shlens, Navdeep Jaitly, and Ian~J. Goodfellow.
\newblock {Adversarial Autoencoders}.
\newblock \emph{CoRR}, abs/1511.05644, 2015.
\newblock URL \url{http://arxiv.org/abs/1511.05644}.

\bibitem[Meehan et~al.(2020)Meehan, Chaudhuri, and Dasgupta]{meehan2020non}
Casey Meehan, Kamalika Chaudhuri, and Sanjoy Dasgupta.
\newblock {A non-parametric test to detect data-copying in generative models}.
\newblock In \emph{International Conference on Artificial Intelligence and
  Statistics}, 2020.

\bibitem[Mondal et~al.(2015)Mondal, Biswas, and Ghosh]{mondal2015high}
Pronoy~K Mondal, Munmun Biswas, and Anil~K Ghosh.
\newblock {On high dimensional two-sample tests based on nearest neighbors}.
\newblock \emph{Journal of Multivariate Analysis}, 141:\penalty0 168--178,
  2015.

\bibitem[M{\"u}ller et~al.(2004)M{\"u}ller, B{\"u}rgi, and
  Dacorogna]{muller2004bootstrapping}
Ulrich~A M{\"u}ller, Roland B{\"u}rgi, and Michel~M Dacorogna.
\newblock {Bootstrapping the economy--a non-parametric method of generating
  consistent future scenarios}.
\newblock 2004.

\bibitem[Nagarajan et~al.(2018)Nagarajan, Raffel, and
  Goodfellow]{nagarajan2018theoretical}
Vaishnavh Nagarajan, Colin Raffel, and Ian~J Goodfellow.
\newblock {Theoretical insights into memorization in GANs}.
\newblock In \emph{Neural Information Processing Systems Workshop}, volume~1,
  2018.

\bibitem[Ni et~al.(2020)Ni, Szpruch, Wiese, Liao, and Xiao]{ni2020conditional}
Hao Ni, Lukasz Szpruch, Magnus Wiese, Shujian Liao, and Baoren Xiao.
\newblock {Conditional Sig-Wasserstein GANs for Time Series Generation}.
\newblock \emph{arXiv preprint arXiv:2006.05421}, 2020.

\bibitem[Pedersen et~al.(2016)Pedersen, Campbell, Christiansen, Cox, Finn,
  Griffin, Hooker, Lightwood, Sonlin, and Suchar]{pedersen2016economic}
Hal Pedersen, Mary~Pat Campbell, Stephan~L Christiansen, Samuel~H Cox, Daniel
  Finn, Ken Griffin, Nigel Hooker, Matthew Lightwood, Stephen~M Sonlin, and
  Chris Suchar.
\newblock {Economic scenario generators: A practical guide}.
\newblock \emph{The Society of Actuaries (July 2016)}, 2016.

\bibitem[Pfeifer and Ragulina(2018)]{pfeifer2018generating}
Dietmar Pfeifer and Olena Ragulina.
\newblock {Generating VaR scenarios under Solvency II with product beta
  distributions}.
\newblock \emph{Risks}, 6\penalty0 (4):\penalty0 122, 2018.

\bibitem[Pritsker(2006)]{pritsker2006hidden}
Matthew Pritsker.
\newblock {The hidden dangers of historical simulation}.
\newblock \emph{Journal of Banking \& Finance}, 30\penalty0 (2):\penalty0
  561--582, 2006.

\bibitem[Radhakrishnan et~al.(2018)Radhakrishnan, Yang, Belkin, and
  Uhler]{radhakrishnan2018memorization}
Adityanarayanan Radhakrishnan, Karren Yang, Mikhail Belkin, and Caroline Uhler.
\newblock {Memorization in overparameterized autoencoders}.
\newblock \emph{arXiv preprint arXiv:1810.10333}, 2018.

\bibitem[Sandstr{\"o}m(2016)]{sandstrom2016handbook}
Arne Sandstr{\"o}m.
\newblock \emph{{Handbook of solvency for actuaries and risk managers: theory
  and practice}}.
\newblock CRC press, 2016.

\bibitem[Scherer and Stahl(2021)]{scherer2021standard}
Matthias Scherer and Gerhard Stahl.
\newblock {The standard formula of Solvency II: a critical discussion}.
\newblock \emph{European Actuarial Journal}, 11:\penalty0 3--20, 2021.

\bibitem[Schilling(1986)]{schilling1986multivariate}
Mark~F Schilling.
\newblock {Multivariate two-sample tests based on nearest neighbors}.
\newblock \emph{Journal of the American Statistical Association}, 81\penalty0
  (395):\penalty0 799--806, 1986.

\bibitem[Shedari(2016)]{shedari2016solvency}
Shahrok Shedari.
\newblock \emph{{Solvency II. A comparison of the standard model with internal
  models to calculate the Solvency Capital Requirements (SCR)}}.
\newblock GRIN Verlag, 2016.

\bibitem[Tobj{\"o}rk(2021)]{tobjork2021value}
David Tobj{\"o}rk.
\newblock {Value at Risk Estimation with Generative Adversarial Networks}.
\newblock 2021.

\bibitem[van~den Burg and Williams(2021)]{van2021memorization}
Gerrit van~den Burg and Chris Williams.
\newblock {On memorization in probabilistic deep generative models}.
\newblock \emph{Advances in Neural Information Processing Systems},
  34:\penalty0 27916--27928, 2021.

\bibitem[van~der Burgt(2019)]{van2019generalAI}
Joost van~der Burgt.
\newblock {General principles for the use of Artificial Intelligence in the
  financial sector}, 2019.
\newblock URL
  \url{https://www.dnb.nl/media/voffsric/general-principles-for-the-use-of-artificial-intelligence-in-the-financial-sector.pdf}.
\newblock [accessed on 2023/10/20].

\bibitem[Varnell(2011)]{varnell2011economic}
EM~Varnell.
\newblock {Economic scenario generators and Solvency II}.
\newblock \emph{British Actuarial Journal}, 16\penalty0 (1):\penalty0 121--159,
  2011.

\bibitem[Wei et~al.(2024)Wei, Zhang, Zhang, Ding, Chen, Ong, Zhang, and
  Xiang]{wei2024memorization}
Jiaheng Wei, Yanjun Zhang, Leo~Yu Zhang, Ming Ding, Chao Chen, Kok-Leong Ong,
  Jun Zhang, and Yang Xiang.
\newblock {Memorization in deep learning: A survey}.
\newblock \emph{arXiv preprint arXiv:2406.03880}, 2024.

\bibitem[Weiss(1960)]{weiss1960two}
Lionel Weiss.
\newblock {Two-sample tests for multivariate distributions}.
\newblock \emph{The Annals of Mathematical Statistics}, 31\penalty0
  (1):\penalty0 159--164, 1960.

\bibitem[Wiese et~al.(2019)Wiese, Bai, Wood, and Buehler]{wiese2019deep}
Magnus Wiese, Lianjun Bai, Ben Wood, and Hans Buehler.
\newblock {Deep hedging: learning to simulate equity option markets}.
\newblock In \emph{33rd Conference on Neural Information Processing Systems
  (NeurIPS 2019), Vancouver, Canada}, 2019.

\bibitem[Wiese et~al.(2020)Wiese, Knobloch, Korn, and
  Kretschmer]{wiese2020quant}
Magnus Wiese, Robert Knobloch, Ralf Korn, and Peter Kretschmer.
\newblock {Quant GANs: deep generation of financial time series}.
\newblock \emph{Quantitative Finance}, 20\penalty0 (9):\penalty0 1419--1440,
  2020.

\bibitem[Xu et~al.(2018)Xu, Huang, Yuan, Guo, Sun, Wu, and
  Weinberger]{xu2018empirical}
Qiantong Xu, Gao Huang, Yang Yuan, Chuan Guo, Yu~Sun, Felix Wu, and Kilian
  Weinberger.
\newblock {An empirical study on evaluation metrics of generative adversarial
  networks}.
\newblock \emph{arXiv preprint arXiv:1806.07755}, 2018.

\end{thebibliography}

\appendix

\part*{Appendix}

\section{Convergence results }\label{sec:Appendix:-Proof-of}
\begin{thm}
\label{lem:weiss}(Weiss, 1960). Let $(\Omega,\mathcal{F},\ensuremath{P)}$
be a probability space. Let $E_{1},...,E_{M},G_{1},...,G_{N}$ be
independent and identically distributed $d$-variate random variables.
Assume $E_{1}$ has a piecewise continuous and bounded probability
density function $f$. Let $\mu\in(0,1]$ and
\[
D_{m}^{\mu}:=\mu\min_{m^{\prime}\ne m}\left\Vert E_{m}-E_{m^{\prime}}\right\Vert _{2},\quad m=1,...,M
\]
and let $S_{m}$, $m=1,...,M$, be the number of points $G_{1},...,G_{N}$
contained in the open sphere $\{x:\,\left\Vert x-E_{m}\right\Vert _{2}<D_{m}^{\mu}\}$\textup{.}
Then the joint distribution of $S_{i}$, $S_{l}$ is the same as the
joint distribution of $S_{i^{\prime}}$ and $S_{l^{\prime}}$ for
any $i\neq l$ and $i^{\prime}\neq l^{\prime}$. It further holds
for any $\alpha>0$ that
\begin{equation}
\lim_{M\to\infty,\frac{M}{N}=\alpha}P[S_{1}=s]=Q(s):=\int_{\mathbb{R}^{d}}\frac{\mu^{-d}\alpha}{\left(1+\mu^{-d}\alpha\right)^{s+1}}f(x)dx,\quad s\in\N_{0}
\end{equation}
and
\[
\lim_{M\to\infty,\frac{M}{N}=\alpha}P[S_{1}=s_{1}\cap S_{2}=s_{2}]=Q(s_{1})Q(s_{2}),\quad s_{1},s_{2}\in\mathbb{N}_{0}.
\]
\end{thm}
\begin{proof}
Lionel Weiss proved the Theorem for $\mu=\frac{1}{2}$ in \citet{weiss1960two},
but his proof works exactly the same for any $\mu\in(0,1]$.
\end{proof}
\begin{proof}
Proof of Theorem \ref{thm:MR_Konvergenz}: Let $\mu:=\rho^{\frac{1}{d}}>0$.
Let $Q$, $D_{m}^{\mu}$, $S_{m}$ as in Theorem \ref{lem:weiss}.
Denote by $f$ the density of $E_{1}$. It holds that
\begin{align*}
 Q(0) & =\frac{\alpha}{\mu^{d}+\alpha}\underbrace{\int_{\mathbb{R}^{d}}f(x)dx}_{=1}=\frac{\alpha}{\rho+\alpha}<\infty.
\end{align*}
Define $R_{m}$ as in Equation (\ref{eq:Rm}) then $\rho^{\frac{1}{d}}R_{m}=D_{m}^{\mu}$
by the definition of $\mu$. Define
\[
Z_{m}:=\mathbf{1}_{[0,\rho^{\frac{1}{d}}R_{m})}\big(\min_{n=1,...,N}\:\left\Vert G_{n}-E_{m}\right\Vert _{2}\big),\quad m=1,...,M.
\]
Then $\Pi_{M,N}^{\rho}=\frac{1}{M}\sum_{m=1}^{M}Z_{m}$ and
\begin{align*}
\mathbf{E}\left[Z_{m}\right] & =P\Big[\min_{n=1,...,N}\:\left\Vert G_{n}-E_{m}\right\Vert _{2}<\rho^{\frac{1}{d}}R_{m}\Big]\\
 & =1-P\Big[\forall n:\left\Vert G_{n}-E_{m}\right\Vert _{2}\ge\rho^{\frac{1}{d}}R_{m}\Big]\\
 & =1-P[S_{m}=0]\\
 & =1-P[S_{1}=0].
\end{align*}
It follows by Theorem \ref{lem:weiss} that
\begin{equation}
\lim_{M\to\infty,\frac{M}{N}=\alpha}\mathbf{E}\left[Z_{m}\right]=1-Q(0)=\frac{\rho}{\rho+\alpha}.\label{eq:Expectation_X_i}
\end{equation}
As $Z_{m}^{2}=Z_{m}$, it follows that
\[
\lim_{M\to\infty,\frac{M}{N}=\alpha}Var(Z_{m})=1-Q(0)-(1-Q(0))^{2}=\frac{\rho\alpha}{(\rho+\alpha)^{2}}<\infty,\quad m=1,2,....
\]
Let $i\neq l$. Then
\begin{align*}
Cov\left(Z_{i},Z_{l}\right) & =\mathbf{E}[Z_{i}Z_{l}]-\mathbf{E}[Z_{i}]\mathbf{E}[Z_{l}]\\
 & =P\left(\min_{n=1,...,N}\:\left\Vert G_{n}-E_{i}\right\Vert _{2}<\rho^{\frac{1}{d}}R_{i}\cap\min_{n=1,...,N}\:\left\Vert G_{n}-E_{l}\right\Vert _{2}<\rho^{\frac{1}{d}}R_{l}\right)-(\mathbf{E}[Z_{1}])^{2}\\
 & =1-P[S_{i}=0\cup S_{l}=0]-(\mathbf{E}[Z_{1}])^{2}\\
 & =1-P[S_{1}=0\cup S_{2}=0]-(\mathbf{E}[Z_{1}])^{2}\\
 & =1-P[S_{1}=0]-P[S_{2}=0]+P[S_{1}=0\cap S_{2}=0]-(\mathbf{E}[Z_{1}])^{2}.
\end{align*}
Therefore we have that
\[
\lim_{M\to\infty,\frac{M}{N}=\alpha}Cov\left(Z_{i},Z_{l}\right)=1-Q(0)-Q(0)+Q^{2}(0)-(1-Q(0))^{2}=0.
\]
By Bienaymé's identity it follows that
\begin{align*}
\mathbf{E}\left[\left|\Pi_{M,N}^{\rho}-\mathbf{E}[Z_{1}]\right|^{2}\right] & =Var\left(\frac{1}{M}\sum_{i=1}^{M}Z_{i}\right)\\
 & =\frac{1}{M^{2}}\left(\sum_{i=1}^{M}Var\left(Z_{i}\right)+\sum_{i\neq l}Cov\left(Z_{i},Z_{l}\right)\right)\\
 & =\frac{1}{M^{2}}\left(MVar\left(Z_{1}\right)+\left(M^{2}-M\right)Cov\left(Z_{1},Z_{2}\right)\right).
\end{align*}
Thus, 
\begin{equation}
\lim_{M\to\infty,\frac{M}{N}=\alpha}\mathbf{E}\left[\left|\Pi_{M,N}^{\rho}-\mathbf{E}[Z_{1}]\right|^{2}\right]=0.\label{eq:mr_Ez}
\end{equation}
Further, by the Minkowski inequality, it holds that
\begin{align*}
\sqrt{\mathbf{E}\left[\left|\Pi_{M,N}^{\rho}-\frac{\rho}{\rho+\alpha}\right|^{2}\right]} & \leq\sqrt{\mathbf{E}\left[\left|\Pi_{M,N}^{\rho}-\mathbf{E}[Z_{1}]\right|^{2}\right]}+\sqrt{\mathbf{E}\left[\left|\mathbf{E}[Z_{1}]-\frac{\rho}{\rho+\alpha}\right|^{2}\right]}.
\end{align*}
Apply Equation (\ref{eq:Expectation_X_i}) and Equation (\ref{eq:mr_Ez})
to conclude.
\end{proof}

\section{Technical details:
dependent data}\label{sec:Technical-details:-dependent}

Let $M,N\in\mathbb{N}$. Assume a business year has $252$ days and
a month consists of $21$ business days. Let $\delta=\frac{1}{252}$.
Let $W^{1},...,W^{d}$ be independent standard Brownian motions. Let
$\Sigma\in\mathbb{R}^{d\times d}$ be symmetric positive definite.
Let $A$ such that $AA^{\intercal}=\Sigma$. Define 
\[
S_{t}^{i}=S_{0}^{i}\exp\left[\left(r-\frac{1}{2}\sum_{h=1}^{d}A_{ih}^{2}\right)t+\sum_{h=1}^{d}A_{ih}W_{t}^{h}\right],\quad i=1,...,d.
\]
We set $S_{0}^{i}=1$, $r=0$, $\Sigma_{ii}=\sigma^{2}$, $\Sigma_{ij}=\sigma^{2}\gamma$
with $\sigma=0.2$ and correlation $\gamma=0.7$. We consider a $d$-dimensional
time series of daily prices of length $M+N+2w$, i.e.,
\[
S_{0}^{i},...,S_{\delta(M+N+2w-1)}^{i},\quad i=1,...,d.
\]
Based on this time series of $M+N+2w$ daily prices, we will construct
two samples of length $M$ and $N$ of log-returns over a time horizon
of $w$ days: Let
\begin{align*}
E_{m}^{i} & =\log\left(\frac{S_{\delta(m-1+w)}^{i}}{S_{\delta(m-1)}^{i}}\right),\quad i=1,...,d,\quad m=1,....,M
\end{align*}
and
\[
G_{n}^{i}=\log\left(\frac{S_{\delta(M+w+n-1+w)}^{i}}{S_{\delta(M+w+n-1)}^{i}}\right),\quad i=1,...,d,\quad n=1,....,N.
\]
If $w=1$ then $E_{m}^{i}$ correspond to daily log-returns and $E_{m}^{i}$
and $E_{m^{\prime}}^{i}$ for $m\neq m^{\prime}$ are independent.
If $w=21$ then $E_{m}^{i}$ corresponds to monthly log-returns. The
two groups $E_{1},...,E_{M}$ and $G_{1},...,G_{N}$ are independent.
However, in the time-dimension, the points inside a group are not
necessarily independent: Let $m>m^{\prime}.$ If $m\geq w+m^{\prime}$
then $E_{m}^{i}$ and $E_{m^{\prime}}^{i}$ are independent since
the Brownian motions have independent increments, otherwise, they
are correlated. A similar statement holds for $G_{n}^{i}$ and $G_{n^{\prime}}^{i}$.

\section{R Code}\label{sec:R-Code}

The following R code implements the statistic $\Pi_{M,N}^{\rho}$
in the function memorizationRatio() and the statistic $T_{NN1,k}$ in
the function TNN1k(). Feel free to replace the for-loops with the
function \emph{mclapply} from the \emph{parallel} package to parallelize
the code.\\
~

{\footnotesize\#Input: matrix D and row number i.}{\footnotesize\par}

{\footnotesize\#Returns: vector of sorted distances of vector D{[}i,
{]} to rows of matrix D{[}-i, {]}.}{\footnotesize\par}

{\footnotesize disIn = function(D, i)\{}{\footnotesize\par}

{\footnotesize ~~~return(sqrt(rowSums((matrix(D{[}i, {]}, nrow
= (nrow(D)-1), ncol = ncol(D), byrow = T) - D{[}-i, {]})\textasciicircum 2)))}{\footnotesize\par}

{\footnotesize\}}{\footnotesize\par}

~

{\footnotesize\#Input: matrix D and vector x with length equal to
number of columns of D.}{\footnotesize\par}

{\footnotesize\#Returns: vector of length nrow(D) of sorted distances
of x to rows of D.}{\footnotesize\par}

{\footnotesize disOut = function(x, D)\{}{\footnotesize\par}

{\footnotesize ~~~return(sqrt(rowSums((matrix(x, nrow = nrow(D),
ncol = ncol(D), byrow = T) - D)\textasciicircum 2)))}{\footnotesize\par}

{\footnotesize\}}{\footnotesize\par}

~

{\footnotesize\#Input: vectors x and y and integer k in \{1,...,lenght(x)\}}{\footnotesize\par}

{\footnotesize\#Returns: the number of elements of x which are in
the first k entries of the contacted and sorted vector c(x, y).}{\footnotesize\par}

{\footnotesize numberSameNearestNeigbor = function(x, y, k)\{}{\footnotesize\par}

{\footnotesize ~~~tmp = matrix(c(x, y, rep(1, length(x)), rep(0,
length(y))), ncol = 2)}{\footnotesize\par}

{\footnotesize ~~~return(sum(tmp{[}order(tmp{[}, 1{]}), {]}{[}1:k,
2{]}))}{\footnotesize\par}

{\footnotesize\}}{\footnotesize\par}

~

{\footnotesize\#Input: M x d matrix of empirical data, N x d matrix
of generated data, rho in (0,1{]}}{\footnotesize\par}

{\footnotesize\#Returns: memorization ratio}{\footnotesize\par}

{\footnotesize memorizationRatio = function(E, G, rho)\{}{\footnotesize\par}

{\footnotesize ~~~d = ncol(E)}{\footnotesize\par}

{\footnotesize ~~~M = nrow(E)}{\footnotesize\par}

{\footnotesize ~~~N = nrow(M)}{\footnotesize\par}

{\footnotesize ~~~MR = 0}{\footnotesize\par}

{\footnotesize ~~~for(m in 1:M)\{}{\footnotesize\par}

{\footnotesize ~~~~~~myDisOut = disOut(E{[}m,{]}, G) \#Distance
of Em to generated data points}{\footnotesize\par}

{\footnotesize ~~~~~~myDisIn = disIn(E, m) \#Distance of Em
to other empirical data points}{\footnotesize\par}

{\footnotesize ~~~~~~if(min(myDisOut) < (min(myDisIn) {*} rho\textasciicircum (1
/ d)))\{}{\footnotesize\par}

{\footnotesize ~~~~~~~~~MR = MR + 1}{\footnotesize\par}

{\footnotesize ~~~~~~\}}{\footnotesize\par}

{\footnotesize ~~~\}}{\footnotesize\par}

{\footnotesize ~~~return(MR / M)}{\footnotesize\par}

{\footnotesize\}}{\footnotesize\par}

~

{\footnotesize\#Input: M x d matrix of empirical data, N x d matrix
of generated data, k in \{1, 2,...\}}{\footnotesize\par}

{\footnotesize\#Returns: the statistic TNN1k}{\footnotesize\par}

{\footnotesize TNN1k = function(E, G, k)\{}{\footnotesize\par}

{\footnotesize ~~~d = ncol(E)}{\footnotesize\par}

{\footnotesize ~~~M = nrow(E)}{\footnotesize\par}

{\footnotesize ~~~N = nrow(G)}{\footnotesize\par}

{\footnotesize ~~~TEk = 0}{\footnotesize\par}

{\footnotesize ~~~TGk = 0}{\footnotesize\par}

{\footnotesize ~~~for(m in 1:M)\{}{\footnotesize\par}

{\footnotesize ~~~~~~myDisOut = disOut(E{[}m,{]}, G) \#Distance
of Em to generated data points}{\footnotesize\par}

{\footnotesize ~~~~~~myDisIn = disIn(E, m) \#Distance of Em
to other empirical data points}{\footnotesize\par}

{\footnotesize ~~~~~~TEk = TEk + numberSameNearestNeigbor(myDisIn,
myDisOut, k)}{\footnotesize\par}

{\footnotesize ~~~\}}{\footnotesize\par}

{\footnotesize ~~~for(n in 1:N)\{ }{\footnotesize\par}

{\footnotesize ~~~~~~myDisOut = disOut(G{[}n,{]}, E) \#Distance
of Gn to empirical data points}{\footnotesize\par}

{\footnotesize ~~~~~~myDisIn = disIn(G, n) \#Distance of Gn
to other generated data points}{\footnotesize\par}

{\footnotesize ~~~~~~TGk = TGk + numberSameNearestNeigbor(myDisIn,
myDisOut, k)}{\footnotesize\par}

{\footnotesize ~~~\}}{\footnotesize\par}

{\footnotesize ~~~TEk = TEk / (M {*} k)}{\footnotesize\par}

{\footnotesize ~~~TGk = TGk / (N {*} k)}{\footnotesize\par}

{\footnotesize ~~~return((M {*} abs(TEk - (M - 1) / (M + N - 1))
+ N {*} abs(TGk - (N - 1) / (M + N - 1))) / (N + M))}{\footnotesize\par}

{\footnotesize\}}{\footnotesize\par}

\section{Data}

\begin{table}[H]
\begin{centering}
\begin{tabular}{|c|c|c|c|c|}
\hline 
{\small Year} & {\small Log-returns} & {\small S\&P 500 Jan.} & {\small S\&P 500 Dec.} & {\small Training/ Testing}\tabularnewline
\hline 
\hline 
{\small 1997} & {\small 0.2751} & {\small 737.01} & {\small 970.43} & {\small Training}\tabularnewline
\hline 
{\small 1998} & {\small 0.2317} & {\small 975} & {\small 1229.23} & {\small Training}\tabularnewline
\hline 
{\small 1999} & {\small 0.1793} & {\small 1228.1} & {\small 1469.25} & {\small Training}\tabularnewline
\hline 
{\small 2000} & {\small -0.0973} & {\small 1455.22} & {\small 1320.28} & {\small Training}\tabularnewline
\hline 
{\small 2001} & {\small -0.1113} & {\small 1283.27} & {\small 1148.08} & {\small Training}\tabularnewline
\hline 
{\small 2002} & {\small -0.2719} & {\small 1154.67} & {\small 879.82} & {\small Training}\tabularnewline
\hline 
{\small 2003} & {\small 0.1994} & {\small 909.03} & {\small 1109.64} & {\small Training}\tabularnewline
\hline 
{\small 2004} & {\small 0.0892} & {\small 1108.48} & {\small 1211.92} & {\small Training}\tabularnewline
\hline 
{\small 2005} & {\small 0.0377} & {\small 1202.08} & {\small 1248.29} & {\small Training}\tabularnewline
\hline 
{\small 2006} & {\small 0.1114} & {\small 1268.8} & {\small 1418.3} & {\small Training}\tabularnewline
\hline 
{\small 2007} & {\small 0.0347} & {\small 1418.3} & {\small 1468.36} & {\small Training}\tabularnewline
\hline 
{\small 2008} & {\small -0.4714} & {\small 1447.16} & {\small 903.25} & {\small Training}\tabularnewline
\hline 
{\small 2009} & {\small 0.1796} & {\small 931.8} & {\small 1115.1} & {\small Training}\tabularnewline
\hline 
{\small 2010} & {\small 0.1044} & {\small 1132.99} & {\small 1257.64} & {\small Training}\tabularnewline
\hline 
{\small 2011} & {\small -0.0113} & {\small 1271.89} & {\small 1257.6} & {\small Training}\tabularnewline
\hline 
{\small 2012} & {\small 0.1104} & {\small 1277.06} & {\small 1426.19} & {\small Testing}\tabularnewline
\hline 
{\small 2013} & {\small 0.2342} & {\small 1462.42} & {\small 1848.36} & {\small Testing}\tabularnewline
\hline 
{\small 2014} & {\small 0.1168} & {\small 1831.98} & {\small 2058.9} & {\small Testing}\tabularnewline
\hline 
{\small 2015} & {\small -0.0070} & {\small 2058.2} & {\small 2043.94} & {\small Testing}\tabularnewline
\hline 
{\small 2016} & {\small 0.1065} & {\small 2012.66} & {\small 2238.83} & {\small Testing}\tabularnewline
\hline 
{\small 2017} & {\small 0.1690} & {\small 2257.83} & {\small 2673.61} & {\small Testing}\tabularnewline
\hline 
{\small 2018} & {\small -0.0727} & {\small 2695.81} & {\small 2506.85} & {\small Testing}\tabularnewline
\hline 
{\small 2019} & {\small 0.2524} & {\small 2510.03} & {\small 3230.78} & {\small Testing}\tabularnewline
\hline 
{\small 2020} & {\small 0.1423} & {\small 3257.85} & {\small 3756.07} & {\small Testing}\tabularnewline
\hline 
{\small 2021} & {\small 0.2530} & {\small 3700.65} & {\small 4766.18} & {\small Testing}\tabularnewline
\hline 
{\small 2022} & {\small -0.2226} & {\small 4796.56} & {\small 3839.5} & {\small Testing}\tabularnewline
\hline 
{\small 2023} & {\small 0.2210} & {\small 3824.14} & {\small 4769.83} & {\small Testing}\tabularnewline
\hline 
\end{tabular}
\par\end{centering}
\caption{S\&P 500, yearly log-returns between beginning of January and end
of December for the years 1997 till 2023.}

\end{table}

\end{document}